\documentclass[11pt]{article}

\usepackage[final]{acl}

\usepackage{times}
\usepackage{latexsym}

\usepackage[T1]{fontenc}

\usepackage[utf8]{inputenc}

\usepackage{microtype}
\usepackage{algorithm}
\usepackage{algorithmic}
\usepackage{bibentry}
\usepackage{makecell}
\usepackage{newfloat}
\usepackage{listings}
\usepackage{enumitem}

\usepackage{xcolor}
\usepackage{pifont}
\usepackage{placeins}

\newcommand{\cmark}{\textcolor{green!60!black}{\ding{51}}}  
\newcommand{\xmark}{\textcolor{red!70!black}{\ding{55}}}    
\usepackage{booktabs}
\usepackage{inconsolata}

\usepackage{graphicx}

\floatstyle{ruled}
\newfloat{listing}{tb}{lst}{}
\floatname{listing}{Listing}

\lstdefinestyle{myjson3}{%
basicstyle=\ttfamily\scriptsize,
columns=fullflexible,
backgroundcolor=\color{gray!8},
frame=tb,
framesep=0pt,
rulecolor=\color{blue!30},
breaklines=true,
breakatwhitespace=false,
postbreak=\mbox{\textcolor{red}{\tiny$\hookrightarrow$}\space},
stringstyle=\color{black!60!black},    
keywordstyle=\color{blue!80!black},    
commentstyle=\color{gray!70},          
identifierstyle=\color{purple!70!black}, 
morekeywords={true, false, null},
morestring=[b]",
morestring=[b]',
literate=
*{:}{{{\color{red!70!black}:}}}1
{,}{{{\color{red!70!black},}}}1,
showstringspaces=false,
numberstyle=\tiny\color{gray!70},
numbers=left,
numberstyle=\tiny\color{red!70},
numbersep=5pt,
stepnumber=1,
firstnumber=1,
}

\lstdefinestyle{llmprompt}{%
basicstyle=\ttfamily\normalsize,
columns=fullflexible,
backgroundcolor=\color{gray!5},
frame=none,
framerule=0pt,
framesep=0pt,
breaklines=true,
keywordstyle=\bfseries,
keywordstyle=[1]\color{black!70!black},     
keywordstyle=[2]\color{black!50!black},    
keywordstyle=[3]\color{black!60!black},   
keywordstyle=[4]\color{black!70!black},   
morekeywords=[1]{system},
morekeywords=[2]{user},
morekeywords=[3]{assistant},
morekeywords=[4]{tool},
numbers=left,
numberstyle=\tiny\color{gray!70},
numbersep=5pt,
stepnumber=1,
firstnumber=1,
numberblanklines=true
}

\definecolor{codegray}{gray}{0.95}
\lstdefinestyle{mypython}{
    backgroundcolor=\color{codegray},
    language=Python,
    basicstyle=\ttfamily\footnotesize,
    keywordstyle=\color{blue}\bfseries,
    commentstyle=\color{gray},
    stringstyle=\color{orange},
    showstringspaces=false,
    numbers=left,
    numberstyle=\tiny\color{gray},
    frame=single,
    breaklines=true,
    tabsize=4,
    captionpos=b
}

\lstdefinestyle{plainpython}{
    language=Python,
    basicstyle=\ttfamily\small,
    numbers=left,
    numberstyle=\tiny,
    frame=single,
    breaklines=true,
    showstringspaces=false,
    keywordstyle=,
    commentstyle=,
    stringstyle=,
}

\title{Live API-Bench: 2500+ Live APIs for Testing Multi-Step Tool Calling }

\author{
Benjamin Elder\thanks{Equal contribution.} \\
IBM Research AI \\
\texttt{benjamin.elder@ibm.com}
\And
Anupama Murthi\textsuperscript{*} \\
IBM Research AI \\
\texttt{anupama.murthi@ibm.com}
\And
Jungkoo Kang \\
IBM Research AI \\
\texttt{jungkoo.kang@ibm.com}
\AND
Ankita Rajaram Naik \\
IBM Research AI \\
\texttt{ankita.naik@ibm.com}
\And
Kiran Kate \\
IBM Research AI \\
\texttt{kakate@us.ibm.com}
\And
Kinjal Basu \\
IBM Research AI \\
\texttt{kinjal.basu@ibm.com}
\AND
Danish Contractor \\
IBM Research AI \\
\texttt{danish.contractor@ibm.com}
}
\begin{document}
\maketitle

\begin{abstract}
Large language models (LLMs) increasingly rely on external tools and APIs to execute complex tasks specified in natural language. Evaluating such {\em tool-calling} capabilities in realistic enterprise settings is challenging: APIs are often proprietary, heterogeneous, and difficult to share, limiting reproducible benchmarks. To address this, we introduce {\bf Live API Bench}, a comprehensive benchmark constructed by transforming NL2SQL datasets into interactive API environments. Our pipeline converts SQL queries from BIRD-SQL into executable API sequences across three formulations—{\em SLOT}, {\em SEL}, and {\em REST}—covering minimal general-purpose operations, domain-specific multi-step tasks, and function-oriented RESTful interactions, respectively. The benchmark spans 11 databases with over 2,500 invocable tools, paired with human-authored queries, ground-truth API sequences, and verified final answers. Live API Bench enables systematic evaluation of core challenges in tool use, including error handling, sequential reasoning, parameter generation, response parsing, and robustness across diverse domains. We evaluate 10 LLMs and 4 ReACT agents, observing low task completion rates (7–47\%), which improve modestly to 50\% under interactive agent settings, highlighting substantial scope for improving LLM tool-calling performance. We release all code and data associated with this paper.  
\end{abstract}


\section{Introduction}

Large language models (LLMs) are increasingly deployed in real-world applications where they must interact with external tools and APIs to accomplish complex tasks expressed in natural language~\citep{yao2023react,xu2023rewoodecouplingreasoningobservations}. Such systems are deployed with access to tools that interact with live environments to accomplish tasks, often in response to a user's request expressed in natural language~\citep{yoran2024assistantbenchwebagentssolve,drouin2024workarena,pan2024webcanvas,OSWorld,zheng-etal-2024-webolympus}. This capability—commonly referred to as {\em tool-calling}—is critical in enterprise environments, where business workflows depend on programmatic access to databases, web services, and specialized software systems via standardized API interfaces.\footnote{\url{https://shorturl.at/bpsps}}$^,$\footnote{\url{https://shorturl.at/daoEp}}

\begin{table*}[ht]
\centering
\scriptsize
\setlength{\tabcolsep}{1mm} 

\resizebox{\textwidth}{!}{%
\begin{tabular}{lcccc ccc crr}
\toprule
\textbf{Benchmarks} & 
\multicolumn{4}{c}{\textbf{Core Features}} & 
\multicolumn{3}{c}{\textbf{Deployment}} & 
\multicolumn{3}{c}{\textbf{Evaluation}} \\
\cmidrule(lr){2-4} \cmidrule(lr){5-8} \cmidrule(lr){9-11}
& \textbf{Hand-} & \textbf{Invocable APIs} & \textbf{Nested API} & \textbf{Code Tools} 
& \textbf{Live API} & \textbf{Accessible} & \textbf{Real-World data} 
& \textbf{Ground Truth} & \textbf{\#} & \textbf{\#} \\
& \textbf{Crafted Queries} & \textbf{for Data Creation} & \textbf{Sequences} & \textbf{as Functions} 
& \textbf{endpoints} & \textbf{DB} & \textbf{powering API} 
& \textbf{Sequences} & \textbf{Instances} & \textbf{Tools} \\
\midrule
\textbf{NesTools} \citep{han2024nestools} & \xmark & \xmark & \cmark & \xmark & \xmark & \xmark & \xmark & \cmark & 1000 & 3034\\
\textbf{ToolAce} \citep{liu2024toolace} & \xmark & \xmark & \cmark & \xmark & \xmark & \xmark & \xmark & \cmark & 11300 & 26507\\
\textbf{API-Bank} \citep{li2023api} & \cmark & \cmark & \xmark & \cmark & \xmark & \cmark & \xmark & \xmark & 314 & 73 \\
\textbf{APIBench} \citep{patil2024gorilla} & \xmark & \cmark & \xmark & \cmark & \cmark & \xmark & \xmark & \xmark & 16450 & 1645\\
\textbf{ToolBench} \citep{xu2023toolmanipulationcapabilityopensource} & \xmark & \cmark & \cmark & \xmark & \cmark & \xmark & \xmark & \cmark & 2746 & 232\\
\textbf{RestBench}\citep{song2023restgpt} & \cmark & \cmark & \cmark & \xmark & \cmark & \xmark & \cmark & \cmark & 157 & 94\\
\textbf{ToolQA} \cite{zhuang2023toolqa} & \xmark & \cmark & \xmark & \cmark & \xmark & \cmark & \cmark & \cmark & 1530 & 13\\
\textbf{ToolAlpaca} \citep{tang2023toolalpacageneralizedtoollearning} & \xmark & \cmark & \xmark & \xmark & \cmark & \xmark & \cmark & \cmark & 3938 & 426\\
\textbf{ToolLLM} \citep{qin2024toolllm} & \xmark & \cmark & \xmark & \xmark & \cmark & \xmark & \cmark & \cmark & 126,486 & 16,464\\
\textbf{APIGen} \citep{NEURIPS2024_61cce86d} & \xmark & \cmark & \xmark & \cmark & \cmark & \xmark & \cmark & \cmark & 60,000 & 3673\\
\textbf{NESTFUL}\citep{basu2025nestfulbenchmarkevaluatingllms} & \cmark & \cmark & \cmark & \cmark & \xmark & \xmark & \xmark & \cmark & 1861 & 921\\
\midrule
\textbf{Ours}  & \cmark & \cmark & \cmark & \cmark & \cmark & \cmark & \cmark & \cmark & 2583 & 2513\\
\bottomrule
\end{tabular}%
}
\caption{Comparison of Invocable API-Bench with existing tool-calling benchmarks, organized by \emph{Core Features}, \emph{Deployment}, and \emph{Evaluation}. While many benchmarks use APIs to construct datasets (see \emph{Invocable APIs} column), these APIs are often not accessible during evaluation (see \emph{Live API endpoints} column).}

\label{tab:dataset-comparison}
\end{table*}

Evaluating LLM tool-calling in realistic deployment settings remains an open challenge: real-world APIs are proprietary, heterogeneous, and difficult to share, which hinders reproducibility. Prior benchmarks thus often rely on simulations, limited domains, or small scales (see Section \ref{sec:related}). To approximate some of the deployment challenges while retaining reproducibility, we build our benchmark from NL2SQL datasets. This design offers three advantages: (1) NL2SQL tasks involve large, diverse databases, (2) they include human-authored natural language queries with guaranteed executable interpretations, and (3) SQL queries can be systematically decomposed into semantically equivalent API-call sequences that highlight practical tool-calling difficulties.

Building on these insights, we introduce {\bf Live API Bench}, a comprehensive benchmark for evaluating LLM tool-calling by transforming NL2SQL resources into interactive API environments. Our data transformation pipeline converts NL2SQL queries from BIRD-SQL~\citep{bird}—one of the largest and most diverse permissively licensed\footnote{Licensed under CC BY-SA 4.0} NL2SQL datasets—into executable API sequences across three complementary formulations:

\begin{itemize}[itemsep=0pt, parsep=0pt, topsep=0.5ex, left=1em]
    \item \textbf{SLOT} — A minimal set of general-purpose APIs (e.g., filter, sort, aggregate) that often require multiple sequential invocations with varying parameters.
    \item \textbf{SEL} — An expanded collection combining domain-specific retrieval functions with general-purpose operations, requiring models to select tools and compose multi-step execution plans.
    \item \textbf{REST} — A function-oriented formulation where queries can typically be resolved through careful function selection and parameter specification, modeling RESTful API interactions.
\end{itemize}

Live API Bench provides fully accessible APIs that can be hosted locally, allowing LLMs and agents to interact with live endpoints during evaluation. Additionally, by deriving APIs from executable SQL queries over real-world databases, our benchmark ensures deterministic, verified final answers, addressing common limitations in existing tool-calling benchmarks (see Table \ref{tab:dataset-comparison}). This setup enables systematic evaluation of tool use with realistic challenges including:  

\begin{itemize}[itemsep=0pt, parsep=0pt, topsep=0.5ex, left=1em]
    \item \textbf{Error handling} — APIs may fail or return malformed responses, requiring models to manage failures and timeouts.
    \item \textbf{Sequential tool calls} — Tasks often involve chaining dependent API calls; our dataset includes sequences of up to eight calls.
    \item \textbf{Parameter generation} — Models must generate appropriate arguments, validate inputs, and infer missing information to bridge natural language queries and structured API schemas.
    \item \textbf{Response parsing} — Successful tool use requires interpreting complex outputs, extracting relevant information (frequently from large API outputs), and recognizing mismatches with intended requests.
    \item \textbf{Scale and diversity} — To avoid overfitting, models are evaluated across diverse domains, API types, and query complexities.
\end{itemize}

\noindent \textbf{Contributions.} In summary, this paper makes the following contributions: (1) We develop novel data transformation pipelines that repurpose existing NL2SQL tasks for evaluating LLM tool-calling, (2) We generate three distinct API formulations—{\em SLOT}, {\em SEL}, and {\em REST}—to study different aspects of tool-calling. Using the 11 publicly available NL2SQL databases from the BIRD-SQL dev set~\citep{bird}, we create over 2,500 invocable tools backed by real databases. Each tool is paired with human-authored natural language queries, a ground-truth API sequence, and a verified final answer. To our knowledge, this is the largest publicly available collection of live, invocable APIs with these characteristics. (3) Finally, we evaluate 10 LLMs and 4 ReACT agents~\citep{yao2023react} on these collections, finding extremely low task completion rates (7–47\%) that improve modestly to 50\% when models interact with the live API environment as ReACT agents, highlighting substantial room for improvement in LLM tool-calling capabilities.  We release all code and data associated with this paper.

\section{Related Work} \label{sec:related}



The emergence of powerful LLMs has spurred the development of data generation pipelines that produce training and evaluation datasets for tool-calling tasks. Such datasets typically consist of a collection of APIs, along with natural language queries paired with ground-truth API calls or sequences of calls~\citep{NEURIPS2024_61cce86d, shi2025taskcraft, qin2024toolllm, basu-etal-2024-api, pereira2024apitestgenie}.

Data generation pipelines generally fall into two categories. Some repurpose existing datasets and queries, leveraging pre-existing data for tool-calling tasks~\citep{peng2021revisiting, basu-etal-2024-api}. Others use LLMs to synthesize realistic queries conditioned on API collections~\citep{qin2024toolllm, tang2023toolalpacageneralizedtoollearning,liu2024toolace,pereira2024apitestgenie,basu2025nestfulbenchmarkevaluatingllms, zhong2025complexfuncbenchexploringmultistepconstrained}. Beyond queries, some work also focuses on generating APIs themselves from structured sources such as textbooks or documentation, which can then be used in tool-calling tasks~\citep{liu2025reftool}.

Methods for data generation are often tailored to the function-calling scenario. Some pipelines generate queries that invoke a single function, potentially multiple times~\citep{tang2023toolalpacageneralizedtoollearning, xu2023toolmanipulationcapabilityopensource}, others focus on queries requiring the invocation of multiple functions including sequencing, and nested function calls~\citep{liu2024toolace,berkeley-function-calling-leaderboard,basu2025nestfulbenchmarkevaluatingllms, zhong2025complexfuncbenchexploringmultistepconstrained}.
To assess the planning and task-resolution capabilities of tool-calling LLMs, recent benchmarks that provide live environments for agents to interact with have also been developed \citep{ruan2024toolemu,yao2024tau,Spider2-V,OSWorld,zhou2024webarenarealisticwebenvironment}. 

\begin{figure}[t!]
\centering
\begin{lstlisting}[style=myjson3,caption={Original NL2SQL Instance},captionpos=t, label=lst:intro-ex-sql]
{"input": "Among the schools with the SAT test takers of over 500, please list the schools that are magnet schools or offer a magnet program.",
"dataset_name": "california_schools",
"query": "SELECT T2.School FROM satscores AS T1 INNER JOIN schools AS T2 ON T1.cds = T2.CDSCode WHERE T2.Magnet = 1 AND T1.NumTstTakr > 500",
"gold_answer": ["Millikan High", "Polytechnic High", "Troy High"]}
\end{lstlisting}
\begin{lstlisting}[style=myjson3,caption={SLOT-BIRD: API sequence},captionpos=t, label=lst:intro-ex-slot]
[{"name": "filter_data",
"arguments": {"data_source": "data_0.csv", "key_name": "schools_Magnet",
"value": 1.0, "condition": "equal_to"},
"label": "data_1.csv"},
{"name": "filter_data",
"arguments": {"data_source": "data_1.csv", "key_name": "satscores_NumTstTakr",
"value": 500.0, "condition": "greater_than"},
"label": "data_2.csv"},
{"name": "retrieve_data",
"arguments": {"data_source": "data_2.csv", "key_name": "schools_School",
"distinct": False, "limit": -1},"label": "retrieved_json"}]
\end{lstlisting}
\hfill
\begin{lstlisting}[style=myjson3,caption={SEL-BIRD: API Sequence},captionpos=t, label=lst:intro-ex-sel]
[{"name": "select_data_equal_to",
"arguments": {"data_source": "data_0.csv", "key_name": "schools_Magnet",
"value": 1.0},
"label": "data_1.csv"},
{"name": "select_data_greater_than",
"arguments": {"data_source": "data_1.csv", "key_name": "satscores_NumTstTakr",
"value": 500.0},
"label": "data_2.csv"},
{"name": "get_schools_Schools",
"arguments": {"data_source": "data_2.csv"},
"label": "retrieved_json"}
]
\end{lstlisting}
\begin{lstlisting}[style=myjson3,caption={REST-BIRD: API endpoint},captionpos=t, label=lst:intro-ex-rest]
{"output": [{
"name": "get_schools_magnet_status_v1_bird_...",
"description": "Get schools with specific magnet status and number of test takers",
"arguments": {
"magnet_status": 1,
"num_test_takers": 500 },
"path": "/v1/bird/california_schools/schools_magnet_status"}]}
\end{lstlisting}
\caption{A sample NL2SQL instance along with the ground-truth API sequence for SLOT-BIRD, SEL-BIRD and REST-BIRD. Function, slot and slot value descriptions not shown for ease of presentation. 
}
\label{fig:data-instance}
\end{figure}

While existing work sometimes includes queries and APIs that were created using live APIs available from proprietary sources ~\citep{chen2024octopusondevicelanguagemodel,qin2024toolllm,basu2025nestfulbenchmarkevaluatingllms, berkeley-function-calling-leaderboard,zhong2025complexfuncbenchexploringmultistepconstrained}, such APIs are usually either unavailable for live agents attempting to answer queries at test time or are only available as paid services by the providers hosting the APIs (eg: RapidAPI Hub).\footnote{\url{http://www.rapidapi.com}} In contrast, our work expands on these prior efforts by generating APIs that can be hosted locally either as endpoints or python tools for use by LLMs (and agents). Further, since our method for API construction relies on the executable SQL queries from the original NL2SQL datasets, we're also able to provide a deterministic and verified ground-truth final-answer, which can be lacking in some tool-calling datasets ~\citep{qin2024toolllm, liu2024toolace}.

Recently, benchmarks that assess the capabilities of LLMs on Data Science tasks have been created~\citep{DSbench}. In contrast to our work such benchmarks require models to solve tasks by generating structured query languages or code~\citep{liu2024survey, li2024autokagglemultiagentframeworkautonomous}.

To the best of our knowledge, our work is the first to demonstrate how NL2SQL datasets can be used to study LLM tool-calling by re-purposing them to provide a collection of invocable APIs backed by real-world data and reusing real-world natural language queries available in the original datasets. 

\section{Dataset}

We construct three NL2API datasets—\emph{SLOT}, \emph{SEL}, and \emph{REST}—using 11 databases from the BIRD development set~\citep{bird}. Each dataset includes: (i) OpenAPI specifications for available APIs, (ii) Live API implementations, (iii) natural language (NL) questions paired with ground-truth API invocation sequences (with slot values), and (iv) the corresponding databases. We only retain samples where the generated API sequences produce results identical to the original SQL queries, Figure~\ref{fig:data-instance} (top) shows an example of an NL2SQL instance.


\medskip
\noindent\textbf{BIRD Collection:} BIRD is one of the largest collections of real-world databases paired with crowd-sourced NL queries and their SQL statements. The development set averages 7 tables, 73 columns, and 358K rows per database. Queries are often complex, involving multiple joins, aggregations, comparisons, and matches, which translate into multi-step API sequences with challenging sequencing and nesting. Many queries also require reasoning over domain knowledge, synonyms, and numeric values.

\subsection{SLOT-BIRD (Slot-Filling Version)}

The SLOT-BIRD version of the dataset was constructed by decomposing generic SQL \emph{SELECT} queries into their constituent parts and mapping each part to an API implemented as a Python function (tool). The design of these APIs is motivated by similar collections used in enterprise applications, such as Tableau\footnote{\url{https://help.tableau.com/current/api/rest_api/en-us/REST/rest_api_concepts_filtering_and_sorting.htm}} or Google Analytics.\footnote{\url{https://developers.google.com/analytics/devguides/reporting/data/v1/rest/v1beta/FilterExpression}}

\medskip
\noindent\textbf{Functions:}  
Joins in the SQL query are handled collectively during an initialization step, implemented programmatically at evaluation time. This step produces a single table, which serves as the starting point for a model or agent to access via APIs when answering the query.\footnote{While it would be possible to require models to perform JOINs themselves, our experiments show that they already struggle on this dataset without this added complexity.} The SLOT-BIRD setup focuses on planning a sequence of data manipulation and access operations using seven APIs:  
\begin{verbatim}
    aggregate_data, filter_data,
    group_data_by, retrieve_data, 
    select_unique_values, sort_data, 
    transform_data
\end{verbatim}  
Each function corresponds to a specific part of an SQL query, such as WHERE, ORDER BY, or COUNT.

\medskip
\noindent\textbf{Slot Values:}  
The functions in the SLOT-BIRD API pool have two types of parameters (slots) that must be filled:  
\begin{enumerate}[noitemsep, topsep=0pt, leftmargin=1em, label=\arabic*.]
    \item \emph{Schema-agnostic slots:} Control operations on the data, e.g., the type of condition in a WHERE clause or the sort order (ascending/descending).  
    \item \emph{Column-specific slots:} Specify which columns to use for operations such as filtering, sorting, and selecting after manipulations. These slots are categorical, corresponding to the columns in the joined table from the initialization step. Each column includes a brief description of its contents.  
\end{enumerate}  

When a tool argument must be chosen from the column set, the OpenAPI specification includes an \texttt{enum} of allowed names and their descriptions. Correctly understanding these names and descriptions constitutes the primary challenge for a tool-calling model or agent on this dataset.

\medskip
\noindent\textbf{Ground-Truth API Sequence:}  
In the example in Listing~\ref{lst:intro-ex-slot}, {\tt filter\_data} (Lines 1, 5) and {\tt retrieve\_data} (Line 9) are two functions. The first invocation of {\tt filter\_data} takes two slot values: the first (Line 2) is a domain-specific slot {\tt schools\_Magnet}, and the second (Line 3) is a control operation that executes an \texttt{equal\_to} comparison on the other slot. This is followed by another {\tt filter\_data} call and finally a {\tt retrieve\_data} call. 

In addition to the function name and arguments, a \emph{label} is provided, referencing the output of the tool call. This allows subsequent steps to access previous outputs, since models cannot reliably manipulate large data objects across function calls. As shown in Line 4 of Listing \ref{lst:intro-ex-slot}, we use files to pass payloads and reference them as input arguments. Models are not responsible for reading or writing these files; this is handled automatically by our evaluation framework. For \emph{SLOT-BIRD}, all tools except {\tt retrieve\_data} save the results of their data manipulation operations into CSV files and return the file path as a string.  

Ground-truth API sequences were constructed programmatically for each NL2SQL instance by parsing the SQL query with the Sqlglot Python library\footnote{\url{https://github.com/tobymao/sqlglot}} and mapping nodes in the resulting syntax tree to tools from the SLOT-BIRD API pool. The output of each ground-truth API sequence is equivalent (up to minor reformatting) to executing the original SQL query on the underlying databases. This output gold answer enables measuring the completion rate of a model or agent, i.e., the proportion of times it produces a sequence of tool calls that leads to the correct result.

\begin{table}
\centering
\scriptsize
\setlength{\tabcolsep}{1mm}
\begin{tabular}{lccccc}
\toprule
\textbf{Dataset} & \textbf{\# Queries} & \textbf{\# Tools} & \textbf{\#Tools Avail./} & \textbf{\# Tool Calls/} & \textbf{\# Slots/}\\
 &  &  & \textbf{Query} & \textbf{Query} & \textbf{Query}\\
\midrule
SLOT-BIRD & 665 & 7 & 7 & 2.7 & 3.29 \\
SEL-BIRD & 651 & 1256 & 49 & 2.9 & 0.05 \\
REST-BIRD & 1257 & 1250 & 125 & 1 & 1.38\\     
\bottomrule  
\end{tabular}
\caption{Invocable API-Bench dataset statistics}
\label{tab:dataset-stats}
\end{table}

\subsection{SEL-BIRD (Selection Version)}




The SEL-BIRD variant is derived from SLOT-BIRD by expanding categorical function arguments into separate functions. Specifically, each possible value of a categorical argument is bound to create a new function with one fewer input parameter. 
For example, in the {\em SLOT-BIRD} API pool, the function `{\tt filter\_data}` has a categorical argument `{\tt condition}`, which can take values such as `{\tt equal\_to}` (Listing \ref{lst:intro-ex-slot}: Line~$3$) or `{\tt greater\_than}` (Listing \ref{lst:intro-ex-slot}: Line~$7$). In the {\em SEL-BIRD} pool, these become distinct functions—`{\tt select\_data\_equal\_to}` (Listing \ref{lst:intro-ex-sel}: Line~$1$) and `{\tt select\_data\_greater\_than}` (Listing \ref{lst:intro-ex-sel}: Line~$5$)—which no longer require a `{\tt condition}` argument. This transformation yields a substantially larger set of tools for data manipulation. 

\noindent{\bf Additional Domain-Specific Functions:} The expansion is even greater for data retrieval. Rather than passing a column name as an argument (as in {\em SLOT-BIRD}), the {\em SEL-BIRD} pool provides a dedicated `{\tt get}` function for each column key (Listing \ref{lst:intro-ex-sel}: Line~$9$). Consequently, the available toolset varies across instances, since the columns in the initialized table depend on the JOINs specified in the underlying SQL query.

\subsection{REST-BIRD Version}

{\bf API Design:} The {\em REST-BIRD} dataset extends the expansion strategy of {\em SEL-BIRD} to the extreme, assigning a dedicated REST endpoint to every instance in the NL2SQL dataset. We adopt a RESTful design to leverage meaningful path parameters, resulting in highly specific and interpretable endpoints. For example, in Listing \ref{lst:intro-ex-rest}, the original NL query is mapped directly to a single API endpoint (Line~$2$).  
Since BIRD contains only `SELECT` queries, all requests in REST-BIRD are {\tt GET}; no {\tt POST}, {\tt PUT}, or {\tt DELETE} requests are included.  

\noindent{\bf Data Generation:} {\em REST-BIRD} produces a very large number of database-specific endpoints, with each user query answered by exactly one API call.  
Instead of manually authoring these endpoints, we employ {\tt Mistral-Large}\footnote{\url{https://huggingface.co/mistralai/Mistral-Large-Instruct-2411}} within an agentic pipeline consisting of four stages:  
(i) {\bf Code Generation Agent} — synthesizes FastAPI server code,  
(ii) {\bf De-duplication Agent} — merges functionally equivalent endpoints (e.g., {\tt getEmployees} vs.\ {\tt getAllEmployees}),  
(iii) {\bf API Execution Module}, and  
(iv) {\bf Verifier and Filtering Agent} — ensures that the generated endpoint produces results identical to the original SQL query. Any instances where the results do not match are discarded and excluded from the dataset. Complete details of this component and the overall data generation process are provided in the Appendix.

The pipeline outputs Python-based FastAPI server code, dockerized and deployed as microservices, resulting in hosted, executable REST APIs.

\section{Experiments} \label{sec:expts}

We use our three datasets to answer the following research questions (i) Do models find the automatically generated API collections challenging?, (ii) How are models affected by the size of the API tool set provided for the task of function selection?, (iii) To what extent do models rely on semantic signals from the function name to solve the task?, (iv) How does the performance improve when models are employed as ReACT agents that can interact with live APIs? 

\subsection{Models and Prompt Formats}\label{sec:models-and-prompts}
We experiment with the following models: Llama 3.1-8B-Instruct and Llama 3.3-70b-instruct~\citep{grattafiori2024llama}, Qwen2.5-7b-instruct and Qwen2.5-72b-instruct ~\citep{yang2024qwen2}, DeepSeek-v3~\citep{liu2024deepseek}, GPT4o-2024-08-06~\citep{hurst2024gpt}, Granite 3.1-8b-instruct\footnote{\url{https://huggingface.co/ibm-granite/granite-3.1-8b-instruct}}, as well as Hammer-7b~\citep{lin2024hammer}, and Watt-8b\footnote{\url{https://huggingface.co/watt-ai/watt-tool-8B}} which are specialized tool-calling models. 

For all evaluations on {\em SLOT} and {\em SEL}, we adopt a prompting format similar to that used in NESTful~\citep{basu2025nestfulbenchmarkevaluatingllms}, as these datasets involve multiple nested API sequences. For evaluations on {\em REST}, we follow a prompting format similar to that used in BFCL~\cite{berkeley-function-calling-leaderboard}. GPT4o limits tool names to 64 characters\footnote{\url{https://community.openai.com/t/function-call-description-max-length/529902}}, so we experiment with two settings: providing APIs as `tools'\footnote{\url{https://platform.openai.com/docs/assistants/tools}} or embedding them directly in the prompts. All LLM prompt templates are included in Appendix~E.


Our input data is formatted to ensure a structured and reliable mechanism for referencing tool outputs, a crucial aspect of tool calling, that allows subsequent tool calls to utilize the results of previous execution. This is achieved by assigning a unique variable name to each tool call that ensures that each tool's output can be distinctly identified and referenced. This approach is particularly important and prevents ambiguity when multiple instances of the same tool, with different arguments, appear within the same tool call sequence (e.g., parallel tool call). With the help of unique identifiers for each tool call, we facilitate clear and efficient tool chaining that ensures the tool dependencies are correctly resolved. All models are required to follow a specified format included in the prompt (Appendix E).

\noindent{\bf Output Parsing:} Although models are instructed to produce outputs in JSON format, they often deviate, leading to instruction-alignment errors that reflect genuine deployment failures. Since proper tool-formatting is essential for functional tool calling, these deviations highlight a critical gap in current models' ability to reliably interface with API systems. To ensure fair evaluation while preserving parseable outputs, we employ robust, model-specific output parsers, as described in Appendix~G.

\noindent {\bf ReACT Agents:}  ReACT agents~\citep{yao2023react} use LLMs to plan one tool call at a time, execute the tool and use the observation to plan the next step. We chose to experiment with larger architecture models over smaller ones, as in early experiments even larger models performed poorly when invoked directly as LLMs without an agent.

\subsubsection{Metrics} \label{sec:metrics}
{\bf Intent:} For evaluating all models, we report a {\em position-aware} intent precision, recall, and $F1$ metrics using a ground-truth API sequence as reference. When a tool call needs to be called more than once but with different slot-values, it is important to track each {\em instance} of a function invocation thus, its sequence of invocation.

\noindent{\bf Slots:} Additionally, we report precision, recall and $F1$ metrics for slots - i.e., the arguments for each function. The slot metrics are conditional on the correct ground-truth intent being predicted but slots are not penalized for missing ground-truth intents. Overall performance, should therefore always be studied along with the Intent metrics.  

\noindent{\bf Completion Rate (Compl. Rate):} Finally, we report the completion rate i.e., the proportion of all instances where the the models were able to return the ground-truth response.  

\begin{table}[h]
\centering
\scriptsize
\setlength{\tabcolsep}{1mm}
\begin{tabular}{@{}l|ccc ccc c@{}}
\toprule
\textbf{Model} & \multicolumn{3}{c}{\textbf{Intent}} & \multicolumn{3}{c}{\textbf{Slot}} & \textbf{Compl.} \\
\cmidrule(lr){2-4} \cmidrule(lr){5-7}
 & P & R & F1 & P & R & F1 & \textbf{Rate} \\
\midrule
\multicolumn{8}{c}{\textbf{SLOT-BIRD}} \\
\midrule
DeepSeek-V3 & 0.82 & 0.54 & 0.65 & 0.67 & 0.67 & 0.67 & {\bf 0.07} \\
GPT4o-2024-08-06 (Prompt) & 0.31 & 0.15 & 0.20 & 0.71 & 0.71 & 0.71 & 0.03 \\
GPT4o-2024-08-06 (Tools) & 0.90 & 0.52 & 0.66 & 0.43 & 0.42 & 0.42 & 0.03 \\
Granite-8b-instruct & 0.63 & 0.53 & 0.58 & 0.44 & 0.41 & 0.43 & 0.00 \\
Hammer2.1-7b & 0.88 & 0.35 & 0.50 & 0.67 & 0.67 & 0.67 & 0.03 \\
Llama-3.1-8B-instruct & 0.58 & 0.09 & 0.16 & 0.61 & 0.62 & 0.61 & 0.00 \\
Llama-3.3-70b-instruct & 0.75 & 0.22 & 0.34 & 0.65 & 0.03 & 0.05 & 0.00 \\
Mixtral-8x22B-instruct & 0.78 & 0.63 & 0.70 & 0.63 & 0.62 & 0.62 & 0.02\\
Qwen2.5-7b-instruct & 0.65 & 0.67 & 0.66 & 0.63 & 0.63 & 0.63 & 0.03 \\
Qwen2.5-72b-instruct & 0.80 & 0.63 & {\bf 0.71} & 0.61 & 0.61 & 0.61 & 0.06 \\
Watt-tool-8b & 0.43 & 0.20 & 0.27 & 0.51 & 0.52 & 0.52 & 0.01\\
\midrule
\multicolumn{8}{c}{\textbf{SEL-BIRD}} \\
\midrule
DeepSeek-V3 & {0.44}  & {0.28}  & {0.34}  & {0.45}  & {0.44}  & {0.44}  & {0.09} \\
GPT4o-2024-08-06 (Prompt) & {0.42}  & {0.39}  & {0.4}  & {0.47}  & {0.46}  & {0.46}  & {0.09} \\
GPT4o-2024-08-06 (Tools) & {0.47}  & {0.30}  & {0.36}  & {0.62}  & {0.57}  & {0.59}  & {0.0} \\
Granite-3.1-8b-Inst & {0.05}  & {0.05}  & {0.05}  & {0.21}  & {0.18}  & {0.2}  & {0.0} \\
Hammer2.1-7b & {0.29}  & {0.16}  & {0.21}  & {0.47}  & {0.4}  & {0.43}  & {0.03} \\
Llama-3.1-8B-Inst & {0.11}  & {0.01}  & {0.02}  & {0.28}  & {0.26}  & {0.27}  & {0.0} \\
Llama-3-3-70b-Inst & {0.41}  & {0.11}  & {0.17}  & {0.25}  & {0.01}  & {0.02}  & {0.0} \\
Mixtral-8x22B-Inst & {0.6}  & {0.5}  & {0.55}  & {0.45}  & {0.44}  & {0.44}  & {0.04} \\
Qwen2.5-7b-Inst & {0.16}  & {0.29}  & {0.2}  & {0.39}  & {0.39}  & {0.39}  & {0.04} \\
Qwen2.5-72b-Inst & {0.55}  & {0.48}  & {0.51}  & {0.46}  & {0.46}  & {0.46}  & {\bf 0.16} \\
Watt-tool-8b & {0.46}  & {0.1}  & {0.16}  & {0.43}  & {0.45}  & {0.44}  & {0.01} \\
\midrule
\multicolumn{8}{c}{\textbf{REST-BIRD}} \\
\midrule
DeepSeek-V3 & 0.65 & 0.50 & 0.57 & 0.77 & 0.74 & 0.76 & 0.31 \\
GPT4o-2024-08-06 (Prompt) & 0.82 & 0.54 & 0.65 & 0.79 & 0.79 & 0.79 & 0.38 \\
Granite-8b-instruct & 0.45 & 0.58 & 0.50 & 0.77 & 0.76 & 0.77 & 0.34 \\
Hammer2.1-7b & 0.70 & 0.22 & 0.34 & 0.89 & 0.86 & 0.87 & 0.17 \\
Llama-3.1-8B-instruct & 0.22 & 0.57 & 0.31 & 0.76 & 0.76 & 0.76 & 0.32 \\
Llama-3.3-70b-instruct & 0.57 & 0.67 & 0.61 & 0.76 & 0.76 & 0.76 & 0.42 \\
Mixtral-8x22B-instruct & 0.46 & 0.39 & 0.42 & 0.78 & 0.77 & 0.77 & 0.24 \\
Qwen2.5-7b-instruct & 0.53 & 0.52 & 0.53 & 0.80 & 0.80 & 0.80 & 0.37 \\
Qwen2.5-72b-instruct & 0.66 & 0.65 & {\bf 0.66} & 0.82 & 0.82 & 0.82 & {\bf 0.47} \\
Watt-tool-8b & 0.60 & 0.64 & 0.62 & 0.78 & 0.76 & 0.77 & 0.43 \\
\bottomrule
\end{tabular}
\caption{Direct invocation performance across datasets and models showing precision (P), recall (R), and F1 scores for intent and slot prediction, along with completion rates.}
\label{tab:main-results}
\end{table}

\begin{figure*}[t]
    \centering
    \includegraphics[width=1.0\linewidth]{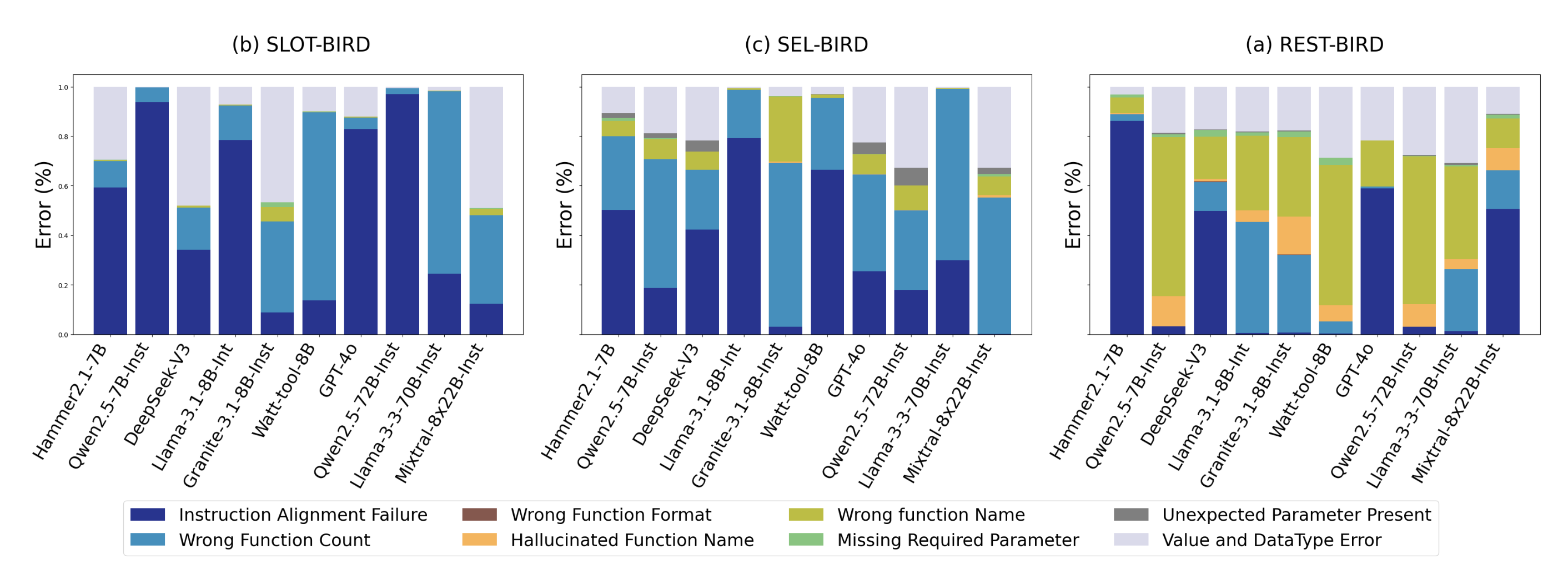}
    \caption{Tool-calling error classification}
    \label{fig:error-analysis}
\end{figure*}

\subsection{Performance of Models} \label{sec:model-performance}

{\bf {\em SLOT-BIRD}}: Recall that this dataset has $7$ general tools and the models need to populate appropriate slot for each invocation of a general tool and also sequence them correctly to accomplish the task. As can be seen from Table \ref{tab:main-results}, this is very challenging task for all models. While the set of APIs to choose from is small, they need to be invoked more than once - this is reflected in the relatively high precision as compared to recall in intent metrics.  We find that Qwen2.5-72b-instruct performs best on the intent selection task though DeepSeek-V3 manages to pair them with the most correct slot-values and reports the highest completion rate ($7$\%).

\noindent{\bf {\em SEL-BIRD}}: This dataset has domain specific tools available that reduce the need to invoke general tools repeatedly. As can be seen from Table \ref{tab:main-results}, models have lower precision and recall as compared to {\em SLOT-BIRD} suggesting that models do not select and sequence right set of tools. As compared to {\em SLOT-BIRD}, all models have lower scores when comparing slots. Notably Granite, has a severe drop in performance on this dataset while the others drop by approx. $0.3$ (absolute) F1 points. As before, based on the completion rates Qwen-72b-instruct and DeepSeek-v3 are the top two performing models.

\noindent {\bf {\em REST-BIRD}:} Finally, in this version of the dataset every query uniquely maps to one invocation and the primary task is of intent selection and then correctly populating the slots. Of the three datasets, this appears to be the one where models have most success with significantly higher intent selection rates (no sequencing needed). While the Qwen-2.5-72b-instruct model continues to be a top-performer, smaller models such as Granite-3.1-8b-instruct and Watt-tool-8b perform better than DeepSeek-v3 and GPT-4o. A closer investigation reveals that for this dataset DeepSeek generates chain-of-thought texts that is relatively harder to standardize for parsing and GPT-4o tends to answer the question instead of picking the function call. Refer to Appendix \ref{sec:appendix-output-parser} for examples from our evaluation.

\noindent{\bf Task Completion Rates:} Models struggle on multi-step function-sequencing tasks, achieving near-zero completion on {\em SLOT-BIRD} and {\em SEL-BIRD}, and only modest rates on {\em REST-BIRD}, far below levels needed for practical deployment. GPT-4o (2024-08-06) reaches 38\% on {\em REST-BIRD}, 9\% on {\em SEL-BIRD}, and 3\% on {\em SLOT-BIRD}. For comparison, GPT-4-0613 evaluated on {\em API-Bank} attains 60.24\%, GPT-4 evaluated on {\em ToolBench} with a depth-first search-based decision tree (DFSDT) prompting strategy achieves 70.4\%, and GPT-4o evaluated on {\em NESTful} reaches 60.0\%, highlighting the relative difficulty of our benchmarks.




We hypothesize that agentic architectures incorporating planning and reflection could substantially increase completion rates. A primary challenge for tool-calling models is accurately interpreting the database schema to populate function slot values. Although schema information is provided via enum descriptions in the tools' OpenAPI specifications and included in the model prompts, effectively leveraging it typically requires multi-step reasoning.

\noindent {\bf Error Analysis} \label{sec:error-analysis}: We employ the BFCL error categories to categorize the errors made by models on the API task~\citep{berkeley-function-calling-leaderboard}. Errors are classified into the first type encountered in the list (See Figure \ref{fig:error-analysis}), so that they are non-overlapping by design. 


Notably, on the {\em SLOT-BIRD} dataset, most failures occur when models struggle to produce outputs in the structured format required for sequencing APIs. On the {\em SEL-BIRD} dataset, this issue is less pronounced, and on {\em REST-BIRD}, other types of errors predominate. This suggests that the complexity of tasks involving multiple function calls and the processing of their outputs challenges models, leading to additional failures such as tool-format issues. While structured decoding could potentially mitigate these errors, prior work has shown that it can also impair model reasoning~\cite{tam-etal-2024-speak}. Together, these patterns highlight the challenges of multi-step reasoning and accurate function selection in complex tool-calling tasks, and our dataset’s characteristics provide a valuable benchmark for studying these nuanced aspects of model behavior.

\subsection{Performance of ReACT agents} \label{sec:react-results}

\begin{table}[h]
\centering
\scriptsize
\setlength{\tabcolsep}{1mm}
\begin{tabular}{@{}l|ccc||ccc@{}}
\toprule
\textbf{Model} & \textbf{Base} & \textbf{Agent} & \textbf{Avg.} & \textbf{OOB} & \textbf{Stuck} & \textbf{Unclassified} \\
 & \textbf{Comp.} & \textbf{Comp.} & \textbf{Loops} &  & & \textbf{Errors} \\
\midrule
\multicolumn{7}{c}{\textbf{SLOT-BIRD}} \\
\midrule
Mixtral-8x22B-instruct & 0.02 & 0.08 & 5.91 & 0.27 & 0.60 & 0.07 \\
Llama-3.3-70b-instruct & 0.00 & 0.06 & 3.28 & 0.05 & 0.32 & 0.59 \\
Qwen2.5-72b-instruct & 0.06 & 0.14 & 5.58 & 0.25 & 0.50 & 0.14 \\
GPT4o-2024-08-06 & 0.03 & {\bf 0.15} & 5.45 & 0.10 & 0.54 & 0.23 \\
\midrule
\multicolumn{7}{c}{\textbf{SEL-BIRD}} \\
\midrule
Mixtral-8x22B-instruct & 0.04 & 0.05 & 6.40 & 0.40 & 0.27 & 0.29 \\
Llama-3.3-70b-instruct & 0.00 & 0.08 & 8.20 & 0.66 & 0.58 & 0.00 \\
Qwen2.5-72b-instruct & 0.16 & {\bf 0.17} & 4.93 & 0.17 & 0.24 & 0.47\\
GPT4o-2024-08-06 & 0.09 & 0.12 & 6.94 & 0.30 & 0.33 & 0.29\\
\midrule
\multicolumn{7}{c}{\textbf{REST-BIRD}} \\
\midrule
Mixtral-8x22B-instruct & 0.24 & 0.32 & 3.16 & 0.00 & 0.02 & 0.65 \\
Llama-3.3-70b-instruct & 0.42 & 0.42 & 3.37 & 0.00 & 0.05 & 0.52 \\
Qwen2.5-72b-instruct & 0.47 & {\bf 0.49} & 3.26 & 0.00 & 0.02 & 0.48 \\
GPT4o-2024-08-06 & 0.38 & {\bf 0.50} & 2.54 & 0.00 & 0.01 & 0.49 \\
\bottomrule
\end{tabular}
\caption{ReACT agent performance across different datasets and models. Completion rates shown as decimals (1.0 = 100\%). OOB = Out-of-budget Errors.}
\label{tab:react-results}
\end{table}

As experimenting with all models as ReACT agents would become very expensive due to think-act-observe (TAO) loops we choose $4$ models from our experiments to serve as ReACT agents. We choose Mixtral-8x22B-instruct which is large mixture-of-expert model that surprisingly performs poorly on our data and wanted to see how the performance changes when it is used as an agent. Additionally, we choose the LLama 3.3-70B-instruct, Qwen2.5-72b-instruct, and GPT4o models for their widespread use. We configure our ReACT agents with a fixed TAO loop budget of $10$ turns.

As can be seen from Table \ref{tab:react-results}, the task completion rates for the agents do not exceed $15$\% and 17\% for the {\em SLOT-BIRD} and {\em SEL-BIRD} datasets respectively, though performance on {\em REST-BIRD} is much higher ($50$\% completion rate by GPT4o). Additionally, agents almost never run out of TAO-loop budget or get stuck repeating the same step on the REST-BIRD dataset but face both problems on the SLOT-BIRD and SEL-BIRD datasets.

\subsubsection{Effect of Obfuscation on tool calling} \label{sec:obfuscation}
To investigate whether models can make use of function descriptions and other meta-data, we repeated experiments shown in Table. \ref{tab:react-results} using non-informative function and argument names. Each API name was replaced by ``func\_N'' where N is a unique integer \citep{paul2025obscuracoder}. 
We found that agent performance dropped slightly (~3\%) for REST-BIRD, but for SLOT-BIRD and SEL-BIRD fell to nearly zero (see Appendix B). 
This is likely due to large amount of descriptive information available in REST-BIRD, with function descriptions for each of the ~1200 APIs. In contrast, tools in SLOT-BIRD and SEL-BIRD are much more generic, and therefore their descriptions were correspondingly more challenging to interpret. 

\subsubsection{Effect of the number of tools available} \label{sec:universe-size}
Since every query in the {\em REST-BIRD} dataset only requires a single API, we experiment by shrinking candidate set size, while ensuring ground-truth intent is always present. The ability of agents to select correct intent does indeed go up as number of choices shrinks. Interestingly, even with $90$\% of the tools removed, the best performing GPT4-o based ReACT agent has a task completion-rate of just $71$\% (see Appendix B). 


\section{Conclusion}
In this paper, we presented Live API Bench, a benchmark that transformed NL2SQL datasets into interactive APIs for evaluating LLM tool-calling capabilities. Spanning 11 databases across diverse domains and including over 2,500 invocable APIs paired with human-authored queries, the benchmark covered three styles—SLOT, SEL, and REST. These tasks exposed models to realistic challenges such as multi-step reasoning, sequential tool calls, parameter generation, and complex response parsing. Our experiments showed that even state-of-the-art LLMs struggled with these tasks, highlighting substantial gaps in current tool-calling abilities and the need for further research on robust API interaction.

This presentation of SLOT-BIRD, SEL-BIRD, and REST-BIRD dataset only included versions packaged as either Python functions or REST APIs. We are currently preparing to release a public benchmark and leaderboard based on an expanded version of this dataset. This benchmark also incorporates RAG tasks, more complex reasoning requirements,  policy/instruction following, and multi-turn dialogues. It will be based on a Model Context Protocol (MCP) implementation of all tools and APIs.

\section{Limitations}

Most NL2SQL datasets, including WikiSQL~\citep{zhong2017seq2sql}, Spider~\citep{yu2018spider}, and the BIRD-SQL dataset~\citep{bird} used in this work, contain only \texttt{SELECT} queries. Data manipulation commands such as \texttt{INSERT}, \texttt{UPDATE}, and \texttt{DELETE} are not included. 
Consequently, our benchmark inherits this limitation: all generated APIs and evaluation tasks are based on \texttt{SELECT} queries. In addition, we note that supporting data-manipulation commands using live APIs would require additional infrastructure, such as containerized database environments or managed services, to ensure safe and deterministic execution, introducing substantial operational complexity.

Furthermore, to guarantee reliable ground-truth verification, the current dataset excludes queries involving nested \texttt{SELECT} statements, \texttt{CASE} expressions, and complex logical compositions of \texttt{WHERE} clauses. Nested and conditional constructs greatly complicate the mapping from SQL syntax trees to tool sequences, as in the SLOT-BIRD formulation, where multiple alternatives make it difficult to track intent and slot metrics. In these cases, the abstract syntax tree is no longer linear but contains hierarchical subqueries or conditional branches, introducing ambiguity in the canonical decomposition into tool calls. By excluding such constructs from this initial version of the dataset, we ensure high-quality, deterministic ground-truth sequences for evaluation. Ongoing work is expanding the scope and complexity of SQL syntax included in this work. 

Due to limitations on computing resources, these exploratory investigations were carried out using only the dev set from the BIRD dataset. The much larger train set has since also been processed and will be included in the public data release, greatly expanding the overall size of the dataset.

\bibliography{aaai2026}
\appendix

\section{Overview}
We include additional experiments by studying the effect of obfuscation on ReACT agents in section \ref{sec:appendix-obfuscation-experiments} and the effect of the size of the universe in Section \ref{sec:appendix-shortlisting-expts}. We describe the artifacts generated by our API generation pipelines in Section \ref{sec:artifacts} and include more details about our data generation pipelines in Section \ref{sec:appendix-data-gen}. We include all LLM prompt templates in Section \ref{sec:appendix_direct_llm_invocation} and agent prompt templates in Section \ref{sec:appendix-agent-experiments}. Finally, we include additional details on output parsing in Section \ref{sec:appendix-output-parser} and how we conducted the error analysis in Section \ref{sec:appendix-bfcl-error-categories}.

\section{Additional Experiments}
\subsection{Effect of Obfuscation}
\label{sec:appendix-obfuscation-experiments}
\subsubsection{Data Construction}
As described in the main paper, our goal was to study the impact of function descriptions and function names. To this end, we obfuscated both API and slot names, replacing them with placeholders such as \texttt{FUNC\_0}, \texttt{FUNC\_1}, ..., \texttt{FUNC\_N} and \texttt{ARG\_1}, \texttt{ARG\_2}, ..., \texttt{ARG\_N}. An example is shown in Fig. \ref{fig:obfuscation}, where an API name and argument gets obfuscated and the obfuscated tool name and slots will be passed to the Agent.

\begin{figure*}
\begin{lstlisting}[
label=lst:unobfuscated, 
caption={Starting API definition},
captionpos=t,
style=myjson3]
"get_free_meal_count_ratio_v1_bird_california_schools_get": {
    "name": "get_free_meal_count_ratio_v1_bird_california_schools_get",
    "description": "Get free meal count ratio for a given county",
    "arguments": {
        "county_name": {
            "type": "string",
            "description": "Name of the county",
            "title": "County Name",
            "name": "ARG_1"
        }
    },
    "path": "/v1/bird/california_schools/free_meal_count_ratio"
}
\end{lstlisting}
\begin{lstlisting}[
label=lst:obfuscated, 
caption={Obfuscated API definition},
captionpos=t,
style=myjson3]
"FUNC_0": {
    "name": "FUNC_0",
    "description": "Get free meal count ratio for a given county",
    "arguments": {
        "ARG_1": {
            "type": "string",
            "description": "Name of the county",
            "title": "County Name",
            "name": "county_name"
        }
    },
    "path": "/v1/bird/california_schools/free_meal_count_ratio"
}
\end{lstlisting}
\caption{\label{fig:obfuscation-example} The pre-obfuscation (Listing \ref{lst:unobfuscated}) and post-obfuscation (Listing \ref{lst:obfuscated}) API specification for an example API endpoint from REST-BIRD. }
\end{figure*}

\begin{figure*}
    \centering
    \includegraphics[width=\textwidth]{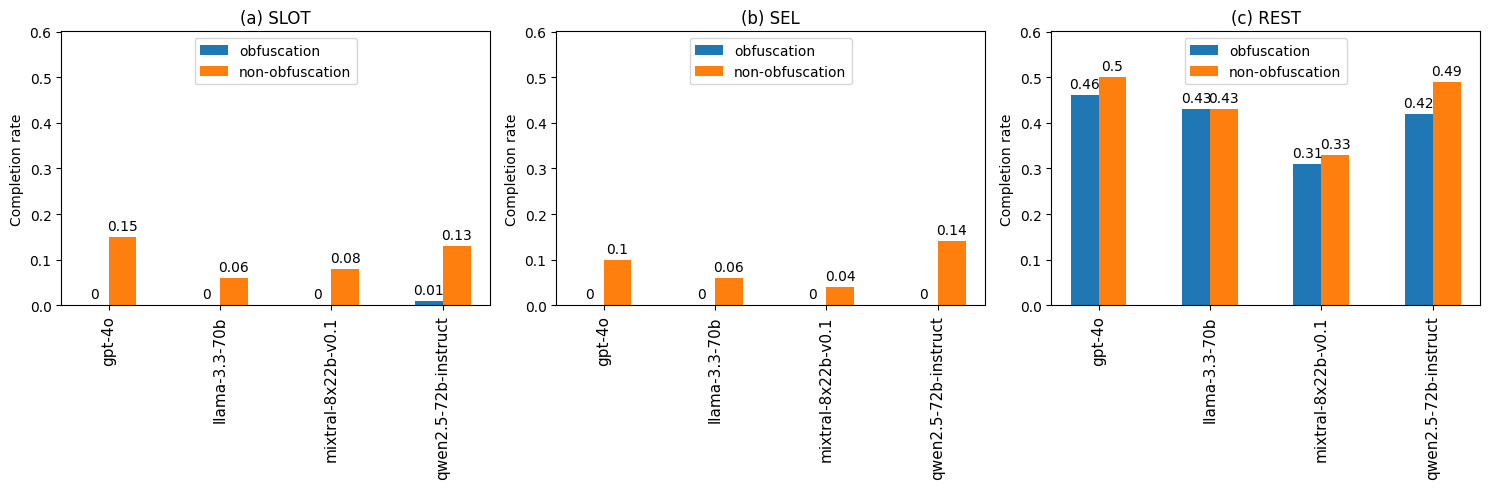}
    \caption{Effect of obfuscation on completion rate}
    \label{fig:obfuscation}
\end{figure*}

\subsubsection{Results}

We re-emphasize a significant contribution of our work: the ability to generate a substantial collection of invocable APIs. We leverage this aspect to investigate the extent to which models can rely on function descriptions and other meta-data when completing tasks with non-informative function names that provide no insight into their intended purpose. To achieve this, we obfuscate each API by assigning a unique integer to the prefix ``func'' \citep{paul2025obscuracoder}. We evaluate performance using the three ReACT agents as before, as they are capable of exploring the environment based on observations (feedback). Unsurprisingly, the performance of models with obfuscated function names is lower across all datasets (Appendix Figure \ref{fig:obfuscation}) with a slight drop in performance on the REST-BIRD dataset but with severe failures on the SLOT-BIRD and SEL-BIRD datasets. This result underscores the importance of employing domain-specific vocabularies to construct tools that enhance agents' ability to improve their tool calling performance. 

\begin{table}[ht]
\centering
\scriptsize
\setlength{\tabcolsep}{1mm}
\begin{tabular}{lrrrrr}
\hline

\textbf{Dataset} & \textbf{10\%} & \textbf{25\%} & \textbf{50\%} & \textbf{75\%} & \textbf{100\%} \\
\hline
formula\_1 & 12 & 31 & 63 & 94 & 126 \\
card\_games & 15 & 38 & 77 & 115 & 154 \\
superhero & 10 & 27 & 54 & 81 & 109 \\
codebase\_community & 15 & 39 & 78 & 117 & 156 \\
thrombosis\_prediction & 13 & 33 & 67 & 101 & 135 \\
toxicology & 9 & 23 & 47 & 71 & 95 \\
financial & 9 & 22 & 45 & 68 & 91 \\
california\_schools & 7 & 18 & 37 & 56 & 75 \\
student\_club & 13 & 32 & 65 & 97 & 130 \\
european\_football\_2 & 11 & 29 & 59 & 88 & 118 \\
debit\_card\_specializing & 6 & 15 & 30 & 45 & 61 \\
\hline
\end{tabular}
\caption{Tool distribution percentiles across datasets}
\end{table}
\subsection{Effect of the number of tools available} \label{sec:appendix-shortlisting-expts}

As described in the main paper, we also examine the effect of varying the number of tools provided to the agent. The REST dataset features a large API universe, and in some cases, over 100 tools are passed to the model—making the selection task significantly more challenging.
\subsubsection{Data Construction}
To study the impact of toolset size, we introduce a best-effort shortlisting mechanism. Given a total of N tools, we reduce the candidate set to 10 percent, 25 percent, 50 percent, or 75 percent of the original size to simulate different levels of tool availability.

We employ a best-effort shortlister that always includes the ground-truth tool and supplements it with N - 1 additional (random) tools, resulting in a set of N tools that is then provided to the model.

For this experiment, we used the REACT agent and the prompt described in the previous section.

The below table shows the list of datasets along with the distribution.

\subsubsection{Results}

\begin{figure}
    \centering
    \includegraphics[width=\columnwidth]{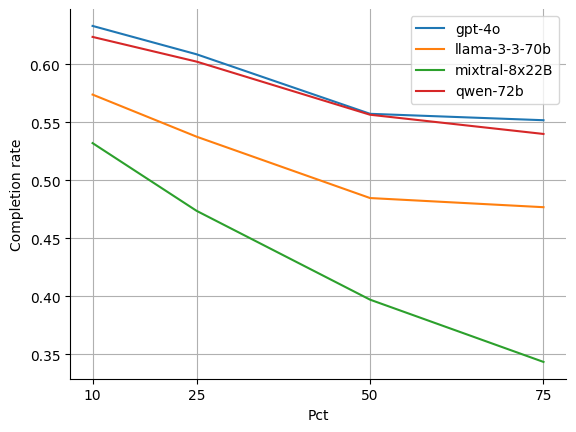}
    \caption{Effect of number of tools (percentage of universe) available on completion rate}
    \label{fig:shortlisting_pct}
\end{figure}
We re-emphasize a significant contribution of our work: the ability to generate a substantial collection of invocable APIs. We leverage this aspect to investigate the extent to which models can rely on function descriptions and other meta-data when completing tasks with non-informative function names that provide no insight into their intended purpose. To achieve this, we obfuscate each API by assigning a unique integer to the prefix ``func'' \citep{paul2025obscuracoder}. We evaluate performance using the three ReACT agents as before, as they are capable of exploring the environment based on observations (feedback). Unsurprisingly, the performance of models with obfuscated function names is lower across all datasets (Appendix Figure \ref{fig:obfuscation}) with a slight drop in performance on the REST-BIRD dataset but with severe failures on the SLOT-BIRD and SEL-BIRD datasets. This result underscores the importance of employing domain-specific vocabularies to construct tools that enhance agents' ability to improve their tool calling performance.

\subsection{Comparison to NL2SQL}

\begin{table}
\centering
\scriptsize
\setlength{\tabcolsep}{1mm}
\begin{tabular}{@{}l|cc|cc|cc @{}}
\hline
\textbf{Model} & \multicolumn{2}{c|}{\textbf{SLOT-BIRD}} & \multicolumn{2}{c|}{\textbf{SEL-BIRD}} &\multicolumn{2}{c}{\textbf{REST-BIRD}} \\
\hline
 & \textbf{API} & \textbf{SQL} & \textbf{API} & \textbf{SQL} & \textbf{API} & \textbf{SQL} \\
\hline
Llama-3.1-8B-instruct & 0.00 & 0.35 & {0.0} & {0.35}& {0.32} & 0.24\\
Llama-3.3-70b-instruct & 0.00 & {\bf 0.45} & {0.0} & {\bf 0.45}& {0.42} & {\bf 0.35}\\
Granite-8b-instruct & 0.00 & 0.13 & {0.0} & {0.13} & 0.34 & 0.09\\
Hammer2.1-7b & 0.03 & -- & {0.03} & {--} & 0.17 & {--}\\
Watt-tool-8b & 0.01 & -- & {0.01} & {--} & 0.43 & {--}\\
Mixtral-8x22B-instruct & 0.02 & 0.29 & {0.04} & {0.29} & 0.24 & 0.22\\
DeepSeek-V3 & {\bf 0.07} & 0.40 & {0.09} & {0.40}& 0.31 & 0.29\\
GPT4o-2024-08-06 (Prompt) & 0.03 & 0.39 & {0.09} & {0.39}& 0.38 & 0.29\\
GPT4o-2024-08-06 (Tools) & 0.03 & -- & {0.0} & {--}& {--} & {--}\\
Qwen2.5-7b-instruct & 0.03 & -- & {0.04} & {--} & 0.37 & --\\
Qwen2.5-72b-instruct & 0.06 & 0.38 & {\bf 0.16} & {0.38}& {\bf 0.47} & 0.27\\ 
\hline
\end{tabular}
\caption{SQL Compl. Rate refers to the task completion rates when models that support SQL generation are used directly to query the database.}
\label{tab:sql-comparison}
\end{table}

Since most LLMs have also been trained to generate SQL queries; we also report the baseline task completion rate when models are prompted to function as SQL-querying systems with direct access to the original database. Interestingly, except on the REST-BIRD dataset, we find that most models report higher completion rates when tasked with generating SQL as opposed to invoking functions. 

\section{Artifacts}
\label{sec:artifacts}
Each data generation pipeline produces the following three artifacts:
\begin{enumerate}
    \item OpenAPI tool/API specifications: These specifications include both the names and parameters needed to invoke the tools, as well as descriptions with the semantic content needed to understand the purpose and behavior of the tools. 
    \item Invocable API/function implementation: The SLOT-BIRD and SEL-BIRD tools are provided as python functions which can be invoked programmatically or bound to a function calling LLM. The REST-BIRD APIs are provided as endpoints in a FastAPI server which has been containerized with Docker for easy deployment. 
    \item Evaluation set: The natural language, tool/API sequence pairs that form the evaluation set can be used to benchmark tool-calling LLMs or Agents. 
\end{enumerate}

\subsection{OpenAPI Functions / Tools} \label{sec:appendix-openapi-functions}

The OpenAPI tool/API specifications include the intent (API) name, description, and input and output parameters. For each parameter, a name, description, and data type is provided. The subset of input parameters that are required in each specification is also provided. For experiments with tool calling LLMs or agents, these specifications can be provided in the model prompt. The model must understand the specifications in order to successfully call the tools. 

Below are some sample tool specifications from each of the three data generation pipelines. 

{\bf SLOT-BIRD}: The `sort\_data' tool from the SLOT-BIRD dataset includes a required `key\_name' parameter which takes values from a domain-specific enum. The possible values and their descriptions are provided as part of the tool specification (see Listing \ref{lst:slot-bird-ex}).

\begin{figure*}[t]
\centering
\begin{lstlisting}[
  caption={SLOT-BIRD APIs: sort\_data tool. },
  label={lst:slot-bird-ex},
  captionpos=t,
  style=myjson3]
{"name": "sort_data", 
"description": "Sort data by the values associated with the chosen key='key_name' If the input data is list-like, returns the sorted list. If the input data is tabular, returns the table with rows sorted by the values in column 'key_name'. If the data is grouped tables, then sort the groups by the value in 'key_name'", 
"parameters": {
    "properties": {
        "data_source": {
            "description": "The location of the data file in csv format.", "schema": {"type": "string"}
        }, 
        "key_name": {
            "description": "name of key to sort by:
            * `member_member_id` - unique id of member
            * `member_first_name` - member's first name
            ...", "schema": {"type": "string", "enum": ["member_member_id", "member_first_name", ...]}
        }, 
        "ascending": {
            "description": "whether to sort by ascending order", "schema": {"type": "boolean"}}}, 
    "required": ["data_source", "key_name", "ascending"], 
    "type": "object"}, 
"output_parameters": {
    "properties": {
        "output_0": {
            "description": "The path to a csv file containing  data sorted by chosen key", "type": "string"
        }
    }
}}
\end{lstlisting}
\end{figure*}

{\bf SEL-BIRD}: The `sort\_data\_ascending' tool from the SEL-BIRD dataset (see Listing \ref{lst:sel-bird-ex}) is equivalent to `sort\_data' in SLOT-BIRD, with the `ascending' argument fixed to True. There is also a corresponding `sort\_data\_descending' tool. 

\begin{figure*}[t]
\centering
\begin{lstlisting}[
  caption={SEL-BIRD APIs: sort\_data\_ascending tool. },
  label={lst:sel-bird-ex},
  captionpos=t,
  style=myjson3]
{"name": "sort_data_ascending", 
"description": "Sort data by the values associated with the chosen key='key_name' If the input data is list-like, returns the sorted list. If the input data is tabular, returns the table with rows sorted by the values in column 'key_name'. If the data is grouped tables, then sort the groups by the value in 'key_name'", 
"parameters": {
    "properties": {
        "data_source": {
            "description": "The location of the data file in csv format.", "schema": {"type": "string"}
        }, 
        "key_name": {
            "description": "name of key to sort by: 
            * `member_member_id` - unique id of member
            * `member_first_name` - member's first name
            ...", "schema": {"type": "string", "enum": ["member_member_id", "member_first_name", ...]}}
            }, 
"required": ["data_source", "key_name", "ascending"], "type": "object"}, "output_parameters": {
    "properties": {
        "output_0": {"description": "The path to a csv file containing  data sorted by chosen key", "type": "string"}
    }
}}
\end{lstlisting}
\end{figure*}

{\bf REST-BIRD}: The REST-BIRD pipeline generates much more specific tools, which were constructed to answer a specific question, rather than perform manipulations on the underlying data,  (see Listing \ref{lst:rest-bird-ex}). 

\begin{figure*}[t]
\centering
\begin{lstlisting}[
  caption={REST-BIRD APIs: get\_zip\_codes example endpoint. },
  label={lst:rest-bird-ex},
  captionpos=t,
  style=myjson3]
{"name": "get_zip_codes_v1_bird_california_schools_zip_codes_get",
"description": "Get zip codes for a given district name and charter school status",
"arguments": {
    "district_name": {
        "type": "string",
        "description": "Name of the district",
        "title": "District Name"
    },
    "charter_school": {
        "type": "integer",
        "description": "Charter school status (1 for yes, 0 for no)",
        "title": "Charter School"
    }
},
"path": "/v1/bird/california_schools/zip_codes"}
\end{lstlisting}
\end{figure*}

\subsection{Evaluation Dataset}

Our data processing pipeline will also provide us with an evaluation set (in JSON format) or the test set will contain the following
\begin{itemize}
    \item Utterance: Natural Language Utterance (from BIRD-SQL dataset)
    \item SQL Query: SQL Query from the BIRD-SQL dataset
    \item Gold Answer: Ground truth obtained by executing the SQL query on the BIRD SQL databases
    \item Output: Dictionary that contains the API (in OpenAI Function format) that will lead to the gold answer
    \item Output from executing the API: The output from executing the API will have the same data as the Gold Answer but it maybe in a different format 
\end{itemize}

During our evaluation, we benchmark various models on the eval-set that we generated using our data generation pipeline.

Refer to Figure 1 for additional details about the eval-set.

\subsection{Invocable APIs (Python Code or Microservices)}

In addition to the previously described components, our data generation pipeline also produces invocable code.

From the REST-BIRD pipeline, we obtain a FastAPI server, which is dockerized and deployed as a microservice.

From the SEQ-BIRD and SEL-BIRD pipelines, we receive a suite of Python tools. These can be invoked by installing the accompanying invocable-api codebase (included with this submission) as a Python library.

\section{Data Generation Pipeline}
\label{sec:appendix-data-gen}

The BIRD SQL dataset provides a rich source of structured information that can be leveraged to construct a high-quality API-centric dataset. Specifically, it includes the following components:

\begin{enumerate} 
    \item Natural language input utterances (questions) 
    \item Corresponding SELECT-SQL queries that yield correct answers 
    \item Real-world databases 
    \item Detailed database schemas, including table names, column names, and descriptions 
    \item Ground truth answers (obtainable by executing the associated SQL queries)     
\end{enumerate}

We get the below information from the BIRD SQL dataset
\begin{lstlisting}[style=myjson3]
{"query": "SELECT `Free Meal Count (K-12)` / `Enrollment (K-12)` FROM frpm WHERE `County Name` = 'Alameda' ORDER BY (CAST(`Free Meal Count (K-12)` AS REAL) / `Enrollment (K-12)`) DESC LIMIT 1",
"input": "What is the highest eligible free rate for K-12 students in the schools in Alameda County?",
"dataset_name": "california_schools",
"gold_answer": [[1.0]]}
\end{lstlisting}

For this work, we utilize the DEV set, which encompasses 11 domains: California Schools, Card Games, Code Base Community, Debit Card Specializing, European Football, Financial, Formula 1, Student Club, Superhero, Thrombosis Prediction, and Toxicology. The BIRD-SQL dataset contains exclusively {\tt SELECT} statements and does not include {\tt UPDATE} or {\tt INSERT} queries. Consequently, the associated tools primarily function as data retrievers, effectively serving as getters.

\subsection{SLOT-BIRD and SEL-BIRD Data Generation Pipelines} \label{sec:appendix-slot-sel-data-gen}

The SLOT-BIRD and SEL-BIRD datasets were constructed by first writing a set of python tools that could perform the same data manipulations as various parts of an SQL SELECT query. The Sqlglot python library was then used to parse each SQL query into an abstract syntax tree. The data generation pipeline then processed the parsed query components in the following order: 

\begin{enumerate}
    \item JOIN statements
    \item WHERE statements
    \item GROUPBY statements
    \item ORDERBY statements
    \item SELECT statements
    \item AGGREGATE statements
\end{enumerate}

{\bf JOIN Statements} These were collectively combined into a single step, carried out by the `initialize\_active\_data' function. This function is not included in the tool specifications provided to the models and agents in the experiments, due to the complex nature of its required arguments. Instead, it is called in the experiment scripts as a setup or data processing step while looping through the data instances. The output of this function is a single table on which the rest of the tool calls will operate. It is saved to a temporary csv file which is pointed to in the model prompts. 

{\bf WHERE Statements} Currently any number of WHERE statements are supported as long as they are connected only by ANDs. The conditions on which the WHERE statements can be performed are: 'equal\_to', 'not\_equal\_to', 'greater\_than', 'less\_than', 'greater\_than\_equal\_to', 'less\_than\_equal\_to', 'contains', and 'like'. Additional conditions such as BETWEEN as well as clauses connected by OR will be included in future versions. 

{\bf GROUPBY Statements} Groupby statements are supported with a parameter to control aggregation type. 

{\bf SELECT Statements} The select statements are handled differently in SLOT-BIRD and SEL-BIRD. In SLOT-BIRD, the `retrieve\_data' function performs this task, returning a subset of the columns of the table provided to it. This function also includes optional parameters: `distinct' which controls whether to only return a list of the distinct elements in the column and `limit' which truncates the returned results. 

In SEL-BIRD, each column is used to create a unique `get\_table\_and\_column\_name' function, which returns only that column. So in this version, selecting N columns requires N tool calls instead of one. The distinct and limit parameters are not available, and instead these are handled by separate functions that must be applied to the output of the `get'. 

A typical example from the student\_club database that illustrates this procedures is shown in Listing \ref{lst:slotsel-simple-example}. 

\begin{figure*}[t]
\begin{lstlisting}[
label=lst:slotsel-simple-example,
caption={Simple example},
captionpos=t,
style=myjson3]
{'query': "SELECT T2.major_name FROM member AS T1 INNER JOIN major AS T2 ON T1.link_to_major = T2.major_id WHERE T1.first_name = 'Angela' AND T1.last_name = 'Sanders'",
 'input': "What's Angela Sanders's major?",
 'gold_answer': 'Business',
 'output': [{'name': 'filter_data',
   'arguments': {'data_source': '$starting_table_var$',
    'key_name': 'member_first_name',
    'value': 'Angela',
    'condition': 'equal_to'},
   'label': 'FILTERED_DF_0'},
  {'name': 'filter_data',
   'arguments': {'data_source': '$FILTERED_DF_0$',
    'key_name': 'member_last_name',
    'value': 'Sanders',
    'condition': 'equal_to'},
   'label': 'FILTERED_DF_1'},
  {'name': 'retrieve_data',
   'arguments': {'data_source': '$FILTERED_DF_1$',
    'key_name': 'major_major_name',
    'distinct': False,
    'limit': -1},
   'label': 'SELECT_COL_0'}],
 'dataset_name': 'student_club',
 'sample_id': 0,
 'initialization_step': {'name': 'initialize_active_data',
  'arguments': {'condition_sequence': [['T1.link_to_major',
     'T2.major_id',
     'INNER']],
   'alias_to_table_dict': {'T1': {'original_table_name': 'member',
     'modified_table_name': 'member'},
    'T2': {'original_table_name': 'major', 'modified_table_name': 'major'}},
   'database_path': '/home/belder/invocable-api-hub/invocable_api_hub/driver/../../db/cache/student_club.sqlite'},
  'label': 'starting_table_var'}}
\end{lstlisting}
\end{figure*}

The example shown in Listing \ref{lst:slotsel-challenging-example} is exceptionally difficult because in order to correctly choose the initial tool in the sequence, a model must not only understand the content of the `Player\_Attributes\_date' column, but also examine the contents to see that it is required to choose a substring from the string value in each cell. This is not evident from the description provided: 
\begin{lstlisting}[style=myjson3]
{'key_name': 'Player_Attributes_date', 'description': 'date', 'dtype': 'string'}
\end{lstlisting}
The need for this transformation can only be inferred by examining the data, which an agent capable of multi-round planning and reflection could accomplish. 

\begin{figure*}
\begin{lstlisting}[
label=lst:slotsel-challenging-example,
caption={Challenging example},
captionpos=t,
style=myjson3]
{
    'dataset_name': 'european_football_2',
     'input': 'Tell the defensive work rate for Kevin Berigaud on 2013/2/22.',
     'query': "SELECT t2.defensive_work_rate FROM Player AS t1 INNER JOIN Player_Attributes AS t2 ON t1.player_fifa_api_id = t2.player_fifa_api_id WHERE SUBSTR(t2.`date`, 1, 10) = '2013-02-22' AND t1.player_name = 'Kevin Berigaud'",
     'gold_answer': 'medium',
     'output': [
         {'name': 'transform_data',
           'arguments': {'data_source': '$starting_table_var$',
            'key_name': 'Player_Attributes_date',
            'operation_type': 'substring',
            'operation_args': {'start_index': 0, 'end_index': 10}},
           'label': 'TRANSFORMED_DF_0'},
          {'name': 'filter_data',
           'arguments': {'data_source': '$TRANSFORMED_DF_0$',
            'key_name': 'Player_Attributes_date',
            'value': '2013-02-22',
            'condition': 'equal_to'},
           'label': 'FILTERED_DF_1'},
          {'name': 'filter_data',
           'arguments': {'data_source': '$FILTERED_DF_1$',
            'key_name': 'Player_player_name',
            'value': 'Kevin Berigaud',
            'condition': 'equal_to'},
           'label': 'FILTERED_DF_2'},
          {'name': 'retrieve_data',
           'arguments': {'data_source': '$FILTERED_DF_2$',
            'key_name': 'Player_Attributes_defensive_work_rate',
            'distinct': False,
            'limit': -1},
           'label': 'SELECT_COL_0'}
       ]
}
\end{lstlisting}
\end{figure*}

\subsubsection{Comparison to Enterprise API Collections} \label{sec:app-enterprise-api}
The SLOT-BIRD API design reflects common patterns found in enterprise analytics platforms. For instance, the Google Analytics Data API uses FilterExpression operations that require specifying field names, operators (equal\_to, greater\_than), and values---directly paralleling our \texttt{filter\_data} function structure.\footnote{\url{https://developers.google.com/analytics/devguides/reporting/data/v1/rest/v1beta/FilterExpression}} Similarly, Tableau's REST API provides generic filtering operations through their Filter class, where users specify field names and comparison operators.\footnote{\url{https://help.tableau.com/current/api/rest_api/en-us/REST/rest_api_concepts_filtering_and_sorting.htm}} These enterprise examples demonstrate that our generic API formulation mirrors real-world tool-calling scenarios rather than being artificially constructed.

Example 1: Tableau APIs, from the filter\_sort\_groups.py example: \footnote{\url{https://github.com/tableau/server-client-python/blob/master/samples/filter_sort_groups.py\#L62}}
\begin{lstlisting}[label=lst:tableau, style=mypython, caption={Tableau Tool}, captionpos=t]
options.filter.add(TSC.Filter(TSC.RequestOptions.Field.Name, TSC.RequestOptions.Operator.Equals, group_name))
\end{lstlisting}

This example has a very similar structure to the filter\_data tool in the SLOT-BIRD dataset, where the field name (key name) and operator (condition) need to be specified.

Similar construction from the Google Analytics APIs:\footnote{\url{https://developers.google.com/analytics/devguides/reporting/data/v1/basics\#python_3}}
\begin{lstlisting}[label=lst:google-analytics, style=mypython, caption={Google Analytics Tool}, captionpos=t]
FilterExpression(
    filter=Filter(
        field_name="browser",
        string_filter=Filter.StringFilter(value="Chrome"),
    )
)
\end{lstlisting}

\subsection{REST-BIRD Data Generation Pipeline} \label{sec:appendix-rest-data-gen}

The 11 dev set domains mentioned above yield a total of 1267 tools and 1250 evaluation data points (test set). Note that our pipeline is easy to extend, and using the train set as well, we will be able to expand our dataset to at least 12,000 tools.

\begin{figure}
    \centering
    \includegraphics[width=\columnwidth]{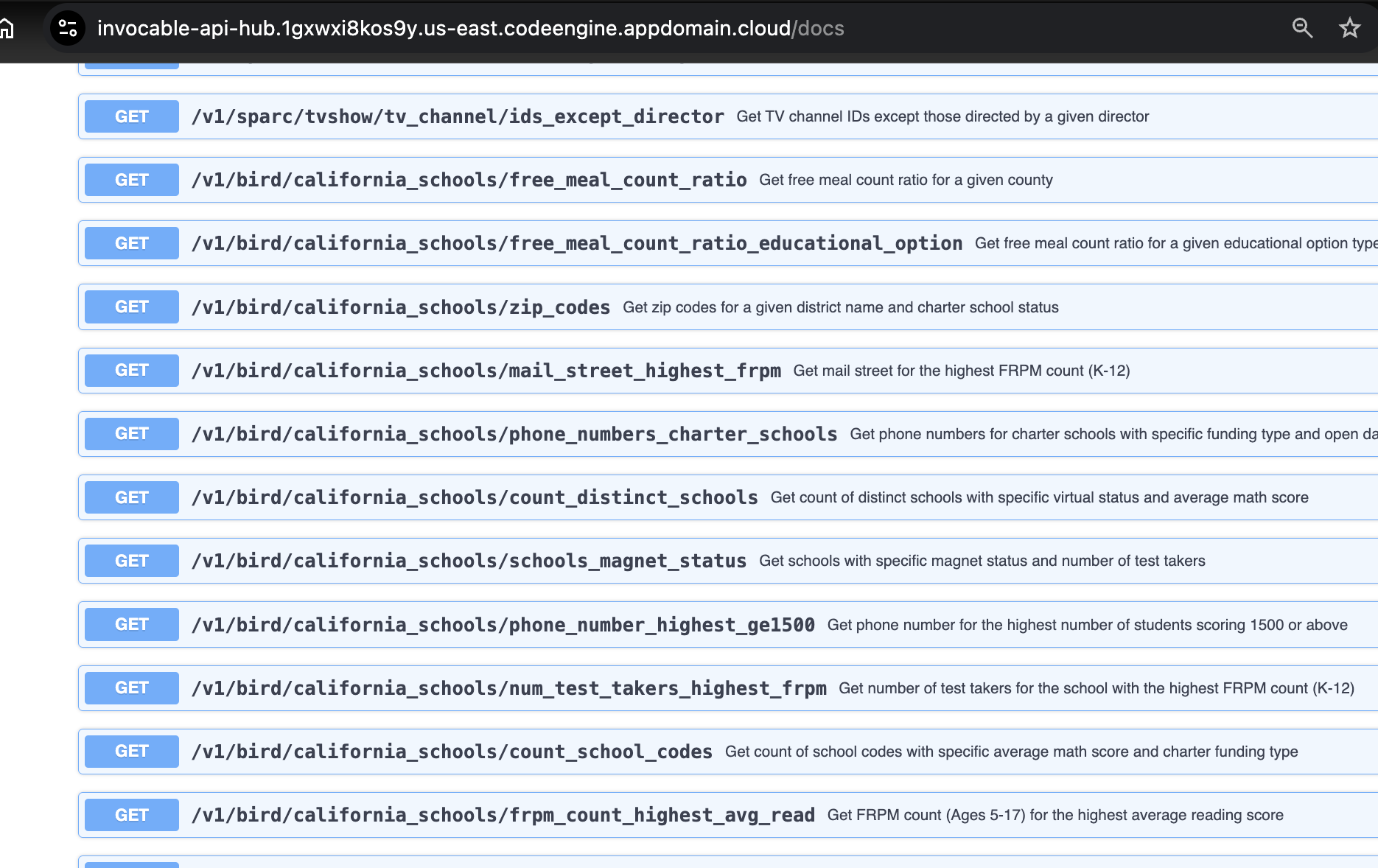}
    \caption{REST API swagger specification, deployed in a server}
    \label{fig:enter-label}
\end{figure}

From the data generation pipeline, we expect the following results
\begin{itemize}
    \item \textbf{Domain-specific API server code} — For each domain (e.g., California Schools, Formula 1), we generate FastAPI server code. This code is then containerized using Docker and deployed to a server, enabling easy access and sharing. We are also relying on the LLM to come up with a description for the API, which is critical as the models rely heavily on the API's description during intent classification. See some sample API implementations in Fig. \ref{fig:enter-label}. 
    
    \item \textbf{Evaluation dataset (JSON)} — A structured JSON file serves as our evaluation dataset, capturing input queries, expected outputs, and metadata. A sample datapoint from the evaluation dataset is shown in Listing \ref{lst:rest-example}

    \item \textbf{API/tool definitions in OpenAPI format} — A list of API endpoints represented in OpenAPI function format, mirroring the generated RESTful endpoints.
\end{itemize}

\begin{figure*}
\centering
\begin{lstlisting}[
label=lst:rest-example,
caption={REST-BIRD Example Datapoint},
captionpos=t,
style=myjson3]
{
    "query": "SELECT `Free Meal Count (K-12)` / `Enrollment (K-12)`\nFROM frpm\nWHERE `County Name` = 'Alameda'\nORDER BY (CAST(`Free Meal Count (K-12)` AS REAL) / `Enrollment (K-12)`) DESC\nLIMIT 1",
    "input": "What is the highest eligible free rate for K-12 students in Alameda County schools?",
    "dataset_name": "california_schools",
    "gold_answer": [[1.0]],
    "output": [{
        "name": "get_free_meal_count_ratio_v1_bird_california_schools_free_meal_count_ratio_get",
        "arguments": {"county_name": "Alameda"},
        "path": "/v1/bird/california_schools/free_meal_count_ratio"
    }],
    "output_after_executing_api": "{\"free_meal_count_ratio\": [1.0]}"
}
\end{lstlisting}
\end{figure*}

In the example  in Listing \ref{lst:rest-example}, we extend the original BIRD-SQL fields by including an additional field, output, which represents the target API endpoint serving as the ground truth. We then execute this API using the FastAPI server code generated by the LLM, and the resulting response is captured in the output\_after\_executing\_api field.

The universe of tools we generate, per dataset (domain) will look like this snippet below. We get the description of the APIs and the description of the SLOTS from the OPEN API Spec (which is auto generated from the FAST API Server Code which we obtain from the Code Generation Agent). 

\begin{figure*}
\centering
\begin{lstlisting}[
label=lst:rest-example-endpoints,
caption={Sample Endpoints},
captionpos=t,
style=myjson3]
    {
        "name": "get_free_meal_count_ratio_educational_option_v1_bird_california_schools",
        "description": "Get free meal count ratio for a given educational option type",
        "arguments": {
            "educational_option_type": {
                "type": "string",
                "description": "Educational option type",
                "title": "Educational Option Type"
            }
        },
        "path": "/v1/bird/california_schools/free_meal_count_ratio_educational_option"
    },
    {
        "name": "get_zip_codes_v1_bird_california_schools_zip_codes_get",
        "description": "Get zip codes for a given district name and charter school status",
        "arguments": {
            "district_name": {
                "type": "string",
                "description": "Name of the district",
                "title": "District Name"
            },
            "charter_school": {
                "type": "integer",
                "description": "Charter school status (1 for yes, 0 for no)",
                "title": "Charter School"
            }
        },
        "path": "/v1/bird/california_schools/zip_codes"
    },
    {
        "name": "get_mail_street_highest_frpm_v1_bird_california_schools_mail_street_highest_frpm_get",
        "description": "Get mail street for the highest FRPM count (K-12)",
        "arguments": {},
        "path": "/v1/bird/california_schools/mail_street_highest_frpm"
    },
\end{lstlisting}
\end{figure*}

Building on this foundation, we investigate the use of large language models (LLMs) to generate realistic, executable REST APIs. To that end, we developed an agentic data generation pipeline composed of several stages as seen below. Some of the stages use a large language models. Some of them use python code to finish a task. 

\begin{enumerate} 
    \item A \textbf{Code Generation Agent}, powered by \texttt{mistral-large}, synthesizes FastAPI server code based on the input utterance, the corresponding SQL query, and the desired API endpoint structure (e.g., \texttt{/v1/bird/formula1/surname?raceid=19}). 
    \item A \textbf{Deduplication Agent} ensures that functionally equivalent endpoints (e.g., \texttt{get\_Employees} vs \texttt{get\_All\_Employees}) are consolidated, avoiding redundant implementations in the generated API. 
    \item The generated API is then executed, and each endpoint is invoked with representative inputs to collect the output. 
    \item \textbf{A Judge Agent}: The output is compared against the expected ground truth results derived from the original SQL query.  We employ a simple judge for this as the output from the REST end point (FAST API Server implementation) maybe different from the output obtained by executing the SQL query.
    \item If the generated output aligns with the expected answer, the API endpoint is considered valid and included in the dataset. Otherwise, the corresponding tool is discarded after a couple of iterative attempts. 
\end{enumerate}

This pipeline enables the creation of a reliable and verifiable dataset of REST APIs grounded in real-world databases and driven by natural language intent.

\subsubsection{Code Generation Agent}

Leveraging an LLM to generate RESTful endpoints offers significant advantages. Key design decisions—such as identifying appropriate query parameters (i.e., input slots), generating descriptive API documentation, and determining standardized output formatting—are effectively automated through the model's capabilities. Furthermore, the data generation pipeline demonstrates strong extensibility, enabling scalable and efficient synthesis of a large number of APIs with minimal manual intervention.

The input to the Code Generation Agent will include a set of SQL queries and the output will be Fast API Server Code.

Refer to the prompt below.
    \begin{lstlisting}[style=llmprompt]
<s>[INST] You are Mixtral Chat, an AI language model developed by Mistral AI. You are a cautious assistant. You carefully follow instructions. You are helpful and harmless and you follow ethical guidelines and promote positive behavior.

I have a list of SQL queries. Take every single query, wrap it into a REST API using Fast API library. Make sure every API/function can handle some query parameters as well.
Definition of query parameter:
Query parameters are added to the end of the URL and are used to filter, sort, or modify the data being returned by the API. For example, a client might want to retrieve a list of products from an e-commerce API, but only those products that are in stock and within a certain price range.

Here is an example:
Consider the follow example query  

SELECT director FROM director WHERE name = 'Pinocchio' 

Look at the where clause in the query above and it has "WHERE name = 'Pinocchio'". I would translate that into a query parameter as seen below. (movie_name)
# Endpoint to get all available years

@app.get("/director", summary="Get director for a given movie")
async def get_director_for_a_movie(movie_name: str = Query(..., description="Name of the movie")):


I am expecting every API to have a query parameter so that it can be generalized nicely.

Here is the list
[
<<TOOLS>>
]
Make sure you use sqlite and cursor to execute the API as well. Make sure the decorator has a nice summary and every API should handle one or more query parameters so that the API is modular and generic. 
Do not make specific APIs. Keep it generic so that it can handle different input values.
Make sure each of the SQL queries above can be mapped to a api end point.

Towards the end, give me the exact API call that would match the eval SQL queries above. Just give me a python list.
Check the total number of API calls you came up with and see if it matches the length of evaluation SQL queries. 

[/INST]
\end{lstlisting}

The above prompt results in LLM-generated FastAPI server code. A snippet of the server code derived from the California Schools domain is shared in Listing \ref{lst:fastapi-server-code}. While this example showcases a few representative endpoints, our pipeline generates at least one endpoint per SQL query in the dataset.

\begin{figure*}[t]
\centering
\begin{lstlisting}[label=lst:fastapi-server-code, style=mypython, caption={FastAPI Endpoint Example}, captionpos=t]

import sqlite3

from fastapi import APIRouter
app = APIRouter()

# Helper function to connect to SQLite and execute queries
def get_db_connection():
    connection = sqlite3.connect('db/california_schools/california_schools.sqlite')
    connection.row_factory = sqlite3.Row  # Enable accessing columns by name
    return connection

# Endpoint to get the free meal count ratio for a given county
@app.get("/v1/california_schools/frpm/free-meal-ratio/{county_name}", summary="Get free meal ratio for a given county")
async def get_free_meal_ratio_for_county(county_name: str):
    connection = get_db_connection()
    cursor = connection.cursor()

    try:
        cursor.execute("SELECT `Free Meal Count (K-12)` / `Enrollment (K-12)` FROM frpm WHERE `County Name` = ? ORDER BY (CAST(`Free Meal Count (K-12)` AS REAL) / `Enrollment (K-12)`) DESC LIMIT 1", (county_name,))
        ratio = cursor.fetchone()[0]
    except Exception as e:
        ratio = ""

    connection.close()
    return ratio

# Endpoint to get the free meal count ratio for continuation schools
@app.get("/v1/california_schools/frpm/free-meal-ratio/continuation-schools", summary="Get free meal ratio for continuation schools")
async def get_free_meal_ratio_for_continuation_schools():
    connection = get_db_connection()
    cursor = connection.cursor()

    try:
        cursor.execute("SELECT `Free Meal Count (Ages 5-17)` / `Enrollment (Ages 5-17)` FROM frpm WHERE `Educational Option Type` = 'Continuation School' AND `Free Meal Count (Ages 5-17)` / `Enrollment (Ages 5-17)` IS NOT NULL ORDER BY `Free Meal Count (Ages 5-17)` / `Enrollment (Ages 5-17)` ASC LIMIT 3")
        ratios = [row[0] for row in cursor.fetchall()]
    except Exception as e:
        ratios = []

    connection.close()
    return ratios
\end{lstlisting}
\end{figure*}

\subsubsection{Judge Agent}

In order to verify the output coming from executing the API vs the output obtained by executing the SQL query, we employ a simple Judge agent (that uses Mixtral 8X22B)

Here is the prompt
\begin{lstlisting}[style=llmprompt]
You are an expert grader. You have ground truth defined below
{ground_truth}
and you have responses from a system defined below.
{response_from_data_gen_pipeline}

Check if the responses and ground truth match properly. It is not a string comparison. You may have to consider parsing the output properly and look for the ground truth in the system response.
In many cases, the ground truth will just have one string but the response may have a long sentence and the ground truth maybe "in the" response. So, carefully parse.
Respond back with a string saying "yes" and "no". "yes" would mean, the ground truth and system responses match up.
I am not expecting the format of the ground truth and the response from the system to match. It is okay if the data is the same but format is different. 
I am not looking for code. Just answer the question, no rambling. Stick to the string output only.
Remember, sometimes the data type may not match but the data between the ground truth and the response may match.
Seperately, tell me the rationale for every decision you made.
Maybe your output can be a comma separate string (decision, rationale)
Note, it is okay if the output format and structure is different from the ground truth. As long as the content matches, we are good.
        
\end{lstlisting}

\section{Direct LLM Invocation Experiments - Prompt Templates}
\label{sec:appendix_direct_llm_invocation}
We conducted experiments using ten different models, including a mix of small- and large-parameter architectures. The following section presents the prompts used in our evaluation. Each prompt instructs the language model to select the appropriate sequence of tools, correctly populate the corresponding slots, and produce a well-formed, machine-parsable JSON output.

\subsection{Prompt Templates for SLOT-BIRD and SEL-BIRD Datasets}
\label{sec:appendix_prompts_slot-sel_datasets}

The SLOT-BIRD and SEL-BIRD prompts used in our non-agentic function-calling experiments are derived from the BFCL prompts, but also include the addition of three ICL examples and an instruction to utilize the initial table created to set up the problem. 

\begin{itemize}
    \item {QUERY}: input natural language question to answer
    \item {FUNCTION\_STR} list of OpenAPI-format specifications for the functions/APIs available to the model
    \item {ICL\_EXAMPLES}: The three ICL examples provided to the model
\end{itemize}

{\bf Llama family of models Prompt (Llama 3.3 70B, Llama 3.1 8B)}
\begin{lstlisting}[
caption={Prompt used for Llama family of models},
captionpos=t,
style=llmprompt]
<|begin_of_text|><|start_header_id|>system<|end_header_id|>

You are an expert in composing functions. You are given a question and a set of possible functions. Based on the question, you will need to make one or more function/tool calls to achieve the purpose. If none of the function can be used, point it out. If the given question lacks the parameters required by the function, also point it out. You should only return the function call in tools call sections.

You have access to the following functions:

{FUNCTION_STR}

 Relevant data is available in the file starting_table_var. 

Here are some examples:
{ICL_EXAMPLES}

Reminder:
- Function calls MUST follow the specified format
- Except the function calls, NO other text MUST be included.

You are a helpful assistant.<|eot_id|><|start_header_id|>user<|end_header_id|>

{QUERY}<|eot_id|><|start_header_id|>assistant<|end_header_id|>
\end{lstlisting}

{\bf watt-tool-8B}
\begin{lstlisting}[
caption={Prompt used for Watt-8B model},
captionpos=t,
style=llmprompt]
<|begin_of_text|><|start_header_id|>system<|end_header_id|>

You are an expert in composing functions. You are given a question and a set of possible functions. Based on the question, you will need to make one or more function/tool calls to achieve the purpose.
If none of the function can be used, point it out. If the given question lacks the parameters required by the function, also point it out.
You should only return the function calls in your response.

If you decide to invoke any of the function(s), If you decide to invoke any of the function, you MUST put it in the JSON format of [{"name": "func_name1", "arguments": {"argument1": "value1", "argument2": "value2"}, "label": "output_label1"}]
You SHOULD NOT include any other text in the response.

At each turn, your should try your best to complete the tasks requested by the user within the current turn. Continue to output functions to call until you have fulfilled the user's request to the best of your ability. Once you have no more functions to call, the system will consider the current turn complete and proceed to the next turn or task.

Here is a list of functions in JSON format that you can invoke.
{FUNCTION_STR} 

 Relevant data is available in the file starting_table_var. 

 Here are some examples: {ICL_EXAMPLES}
<|eot_id|><|start_header_id|>user<|end_header_id|>

{QUERY}<|eot_id|><|start_header_id|>assistant<|end_header_id|>

\end{lstlisting}

{\bf Mixtral 8X22B Prompt}
\begin{lstlisting}[
caption={Prompt used for Mixtral 8x22B model},
captionpos=t,
style=llmprompt]
<|begin_of_text|><|start_header_id|>system<|end_header_id|>

You are an expert in composing functions. You are given a question and a set of possible functions. Based on the question, you will need to make one or more function/tool calls to achieve the purpose. If none of the function can be used, point it out. If the given question lacks the parameters required by the function, also point it out. You should only return the function call in tools call sections.

You have access to the following functions:

{FUNCTION_STR}

 Relevant data is available in the file starting_table_var. 

Here are some examples:
{ICL_EXAMPLES}

Reminder:
- Function calls MUST follow the specified format
- Except the function calls, NO other text MUST be included.

You are a helpful assistant.<|eot_id|><|start_header_id|>user<|end_header_id|>

{QUERY}<|eot_id|><|start_header_id|>assistant<|end_header_id|>
>>> print(prompts['Mixtral-8x22B-Instruct-v0.1'])
[AVAILABLE_TOOLS] {FUNCTION_STR} [/AVAILABLE_TOOLS] 

 Relevant data is available in the file starting_table_var. [INST] Here are some examples:
{ICL_EXAMPLES}

{QUERY}
Generate the output function call strictly in a JSON format. No other text should be included. [/INST]
\end{lstlisting}

{\bf DeepSeek V3 Prompt}
\begin{lstlisting}[
caption={Prompt used for DeepSeek V3 model},
captionpos=t,
style=llmprompt]
<|begin_of_text|><|start_header_id|>system<|end_header_id|>

You are an expert in composing functions. You are given a question and a set of possible functions. Based on the question, you will need to make one or more function/tool calls to achieve the purpose. If none of the function can be used, point it out. If the given question lacks the parameters required by the function, also point it out. You should only return the function call in tools call sections.

You have access to the following functions:

{FUNCTION_STR}

 Relevant data is available in the file starting_table_var. 

Here are some examples:
{ICL_EXAMPLES}

Reminder:
- Function calls MUST follow the specified format
- Except the function calls, NO other text MUST be included.

You are a helpful assistant.<|eot_id|><|start_header_id|>user<|end_header_id|>

{QUERY}<|eot_id|><|start_header_id|>assistant<|end_header_id|>
\end{lstlisting}

{\bf Qwen family of models Prompt (Qwen 2.5 7B, Qwen 2.5 72B)}

\begin{lstlisting}[
caption={Prompt used for Qwen family of models},
captionpos=t,
style=llmprompt]
<|im_start|>system
You are Qwen, created by Alibaba Cloud. You are a helpful assistant.

# Tools

You may call one or more functions to assist with the user query.

You are provided with function signatures within <tools></tools> XML tags:
<tools>
{FUNCTION_STR}
</tools>

For each function call, return a json object with function name and arguments within <tool_call></tool_call> XML tags:
<tool_call>
{"name": <function-name>, "arguments": <args-json-object>, "label": <output-label>}
</tool_call> 

 Relevant data is available in the file starting_table_var. 

 Here are some examples: {ICL_EXAMPLES}
<|im_end|>
<|im_start|>user
{QUERY}<|im_end|>
<|im_start|>assistant
\end{lstlisting}

{\bf Granite 3.1 8B Instruct prompt}

\begin{lstlisting}[
caption={Prompt used for Granite-8B model},
captionpos=t,
style=llmprompt]
<|start_of_role|>system<|end_of_role|>You are a helpful assistant with access to the following function calls. Your task is to produce a sequence of function calls necessary to generate response to the user utterance. Relevant data is available in the file starting_table_var. <|end_of_text|>
\end{lstlisting}

{\bf GPT-4o prompt}

\begin{lstlisting}[
caption={Prompt used for GPT-4o model},
captionpos=t,
style=llmprompt]
You are a helpful assistant with access to the following function calls. Your task is to produce a sequence of function calls necessary to generate response to the user utterance. 

 Relevant data is available in the file starting_table_var. 

 Here are some examples: {ICL_EXAMPLES}
\end{lstlisting}

{\bf Hammer prompt}

\begin{lstlisting}[
caption={Prompt used for Hammer model},
captionpos=t,
style=llmprompt]
<|im_start|>system
You are a helpful assistant.<|im_end|>
<|im_start|>user
[BEGIN OF TASK INSTRUCTION]
You are a tool calling assistant. In order to complete the user's request, you need to select one or more appropriate tools from the following tools and fill in the correct values for the tool parameters. Your specific tasks are:
1. Make one or more function/tool calls to meet the request based on the question.
2. If none of the function can be used, point it out and refuse to answer.
3. If the given question lacks the parameters required by the function, also point it out.

[END OF TASK INSTRUCTION]

[BEGIN OF AVAILABLE TOOLS]
{FUNCTION_STR}
[END OF AVAILABLE TOOLS]

[BEGIN OF FORMAT INSTRUCTION]

The output MUST strictly adhere to the following JSON format, and NO other text MUST be included.
The example format is as follows. Please make sure the parameter type is correct. If no function call is needed, please directly output an empty list '[]'
```
[
    {"name": "func_name1", "arguments": {"argument1": "value1", "argument2": "value2"}, "label": "output_label1"},
    ... (more tool calls as required)
]
```

 Relevant data is available in the file starting_table_var. 

Here are some examples: {ICL_EXAMPLES}

[END OF FORMAT INSTRUCTION]

[BEGIN OF QUERY]
{QUERY}
[END OF QUERY]

<|im_end|>
<|im_start|>assistant
\end{lstlisting}

\subsection{Prompt Templates for REST-BIRD Dataset}
\label{sec:appendix_prompts_rest_datasets}

To the best of our knowledge, for experiments that involve directly querying the LLM using the REST datasets, we have adopted the same prompt format as used in the experiments presented on the BFCL leaderboard \cite{berkeley-function-calling-leaderboard}.

The prompt requires specific components to be filled in. It is important to note that, consistent with the BFCL setup, we do not use in-context learning examples in these experiments involving the REST datasets.

\begin{itemize}
    \item Query (refers to the input utterance)
    \item Tools (refers to the list of functions / api's)
\end{itemize}

{\bf Llama family of models Prompt (Llama 3.3 70B, Llama 3.1 8B, Watt-tools 8B)}
\begin{lstlisting}[
caption={Prompt used for Llama family of models},
captionpos=t,
style=llmprompt]
<|begin_of_text|><|start_header_id|>system<|end_header_id|>

You are an expert in composing functions. You are given a question and a set of possible functions. 
Based on the question, you will need to make one or more function/tool calls to achieve the purpose.
If none of the function can be used, point it out. If the given question lacks the parameters required by the function, also point it out.

You should only return the function calls in your response.
If you decide to invoke any of the function(s), you MUST put it in the format of [func_name1(params_name1=params_value1, params_name2=params_value2...), func_name2(params)]
You SHOULD NOT include any other text in the response.
At each turn, your should try your best to complete the tasks requested by the user within the current turn. 
Continue to output functions to call until you have fulfilled the user's request to the best of your ability. Once you have no more functions to call, the system will consider the current turn complete and proceed to the next turn or task.
Here is a list of functions in JSON format that you can invoke.
{FUNCTION_STR}<|eot_id|><|start_header_id|>user<|end_header_id|>
{QUERY}<|eot_id|><|start_header_id|>assistant<|end_header_id|>
\end{lstlisting}

{\bf Mixtral 8X22B Prompt}
\begin{lstlisting}[
caption={Prompt used for Mixtral 8x22B model},
captionpos=t,
style=llmprompt]
[INST]You are a helpful assistant with access to the following function calls. 
Your task is to produce a sequence of function calls necessary to generate response to the user utterance. 
Use the following function calls as required and return only function "name" with empty "arguments" as a list of dictionary in your response.[/INST]
[AVAILABLE_TOOLS] {FUNCTION_STR} [/AVAILABLE_TOOLS] 

USER: {QUERY} 
ASSISTANT: 
\end{lstlisting}

{\bf DeepSeek V3 Prompt}
\begin{lstlisting}[
caption={Prompt used for DeepSeek V3 model},
captionpos=t,
style=llmprompt]
<|begin_of_sentence|>You are a helpful AI assistant with access to the following tools. 
When a tool is required to answer the user's query, respond with the tool call followed by a JSON list of tools used. 
If a tool does not exist in the provided list of tools, notify the user that you do not have the ability to fulfill the request. 
Use the following function calls as required and return only function ``name'' with empty ``arguments'' as a list of dictionary in your response.Here are the tools. {FUNCTION_STR}<|User|>{QUERY}<|Assistant|>
\end{lstlisting}

{\bf Qwen family of models Prompt (Qwen 2.5 7B, Qwen 2.5 72B)}

\begin{lstlisting}[
caption={Prompt used for Qwen family of models},
captionpos=t,
style=llmprompt]
<|im_start|>system
You are Qwen, created by Alibaba Cloud. You are a helpful assistant.

# Tools

You may call one or more functions to assist with the user query.

You are provided with function signatures within <tools></tools> XML tags:
<tools>
{FUNCTION_STR}
</tools>

For each function call, return a json object with function name and arguments within <tool_call></tool_call> XML tags:
<tool_call>
{"name": <function-name>, "arguments": <args-json-object>}
</tool_call><|im_end|>
<|im_start|>user
{QUERY}<|im_end|>
<|im_start|>assistant
\end{lstlisting}

{\bf Granite 3.1 8B prompt}

\begin{lstlisting}[
caption={Prompt used for Granite 8B model},
captionpos=t,
style=llmprompt]
<|start_of_role|>system<|end_of_role|>
You are a helpful assistant with access to the following function calls. Your task is to produce a single function call necessary to generate response to the user utterance.<|end_of_text|>
<|start_of_role|>system<|end_of_role|>
Knowledge Cutoff Date: April 2024. Today's Date: February 19, 2025. 
You are Granite, developed by IBM. 
You are a helpful AI assistant with access to the following tools. 
When a tool is required to answer the user's query, respond with <|tool_call|> followed by a JSON list of tools used. 
If a tool does not exist in the provided list of tools, notify the user that you do not have the ability to fulfill the request.
<|end_of_text|>
<|start_of_role|>tools<|end_of_role|>
{FUNCTION_STR}
<|start_of_role|>user<|end_of_role|>{QUERY}<|end_of_text|>
<|start_of_role|>assistant<|end_of_role|>
\end{lstlisting}

{\bf GPT-4o prompt}

\begin{lstlisting}[
caption={Prompt used for GPT-4o model},
captionpos=t,
style=llmprompt]
You are a helpful assistant with access to the following function calls. 
Your task is to produce a sequence of function calls necessary to generate response to the user utterance. 
If you decide to invoke any of the function, you MUST put it in the format of [{\"name\": \"func_name1\", \"arguments\": {\"argument1\": \"value1\", \"argument2\": \"value2\"}}].

\n\nAvailable Tools:\n {FUNCTION_STR}
{QUERY}
\end{lstlisting}

{\bf Hammer prompt}

\begin{lstlisting}[
caption={Prompt used for Hammer model},
captionpos=t,
style=llmprompt]
<|im_start|>system
You are a helpful assistant.<|im_end|>
<|im_start|>user
[BEGIN OF TASK INSTRUCTION]
You are a tool calling assistant. In order to complete the user's request, you need to select one or more appropriate tools from the following tools and fill in the correct values for the tool parameters. Your specific tasks are:
1. Make one or more function/tool calls to meet the request based on the question.
2. If none of the function can be used, point it out and refuse to answer.
3. If the given question lacks the parameters required by the function, also point it out.
The following are characters that may interact with you
1. user: Provides query or additional information.
2. tool: Returns the results of the tool calling.

[END OF TASK INSTRUCTION]

[BEGIN OF AVAILABLE TOOLS]
{FUNCTION_STR}
[END OF AVAILABLE TOOLS]

[BEGIN OF FORMAT INSTRUCTION]

The output MUST strictly adhere to the following JSON format, and NO other text MUST be included.
The example format is as follows. Please make sure the parameter type is correct. If no function call is needed, please directly output an empty list '[]'
```
[
    {"name": "func_name1", "arguments": {"argument1": "value1", "argument2": "value2"}},
    ... (more tool calls as required)
]
```

[END OF FORMAT INSTRUCTION]

[BEGIN OF QUERY]
{QUERY}
[END OF QUERY]

<|im_end|>
<|im_start|>assistant
\end{lstlisting}

\section{Agent Experiments - Prompt Templates}
\label{sec:appendix-agent-experiments}
In our experiment, we use a REACT-style agent capable of both tool invocation and self-reflection. The agent operates within a Think-Act-Observe (TAO) loop and is allowed a maximum of 10 iterations. Its objective is to select the appropriate sequence of tools along with their corresponding slot arguments, execute the final API call, and retrieve the resulting response.

The prompt requires the following components to be filled in. 
\begin{itemize}
    \item Query/Input (refers to the input utterance)
    \item Tools (refers to the list of functions / api's)
    \item Previous Runs (the TAO loops from the previous iterations)
\end{itemize}
\subsection{REACT prompt template}
\begin{lstlisting}[style=llmprompt]

Answer the following questions as best you can. 
You have access to the following tools:

{tools}

Respond to the query using the available APIs.

The format you use the API is by specifying 
1) Action: the API function name you would like to call 
2) Action Input: the input parameters of the API call in a json string format. 
The result of the API call will be returned starting with "Observation:". Remember that you should only perform a SINGLE action at a time, do NOT return a list of multiple actions.

Reminder:
1) the only values that should follow "Action:" are: {tool_names}
2) use the following json string format for the API arguments:

Action Input:
{{
    "key_1": "value_1",
    ...
    "key_n": "value_n",
}}

Remember to ALWAYS use the following format:

Thought: you should always think about what to do next
Action: the API function name
Action Input: the input parameters of the API call in json string format
Observation: the return result of the API call. This is what I will provide you with; you do not need to repeat it in your response.
... (this Thought/Action/Action Input/Observation can repeat N times)
Thought: I now know the final answer
Final Answer: the response to the user query

Begin! Remember that your response should never start with "Observation:" since that is what I will provide you with.

Question: {input}

{previousruns}

Thought:{agent_scratchpad}
\end{lstlisting}

\section{Output Parsing and Error Analysis}

\label{sec:appendix-output-parser}

For the SLOT-BIRD and SEL-BIRD experiments, we employed a four stage parsing procedure. First we assume that the input is valid JSON or JSONL format and attempt a json.loads(). If this fails, we attempt to parse using the Python ast library. If this fails as well, we attempt an xml parsing strategy to handle tool calls in xml brackets(<tool\_call/>). Common formatting errors such as prepending text to the JSON body and enclosing the body in extra sets of brackets are handled.

For the REST-BIRD experiments, in most cases, after we receive an output from the model, we use a REGEX-based parser to extract sequences of APIs and their arguments from the model's generated text. However, even with custom parsers in place, some failures persist. We believe these are instruction alignment failures as the model is unable to adhere to the prompt.

Here are some REGEX patterns we look for, across various models
\begin{lstlisting}[style=mypython]
    pattern = r"```json\s*(.*?)\s*```" # Helps parse Deepseek output
    pattern = r"<tool_call>\s*(.*?)\s*</tool_call>" # Hammer/Qwen output 
\end{lstlisting}

\subsection{REST-BIRD analysis on Direct LLM invocation experiments}

\label{sec:appendix-rest-error}
As discussed in the main paper, Qwen2.5-72B-instruct achieves the highest win rate, followed by watt-tool-8b and Llama3-3-70B-instruct. Interestingly, even larger models like DeepSeek-V3, GPT-4o, Mixtral-8x22B show lower win rates, despite having reasonably strong Intent F1 and Slot F1 scores. This drop in win rate is largely due to model's instruction alignment failures.

For our evaluation, we used BFCL prompts. While we employed best-effort parsing, incorporating ICL examples and further prompt engineering could help mitigate some of these issues. That said, we believe it's reasonable to penalize models that fail to follow the prompt format or return poorly structured outputs that are difficult to parse.

Here is an example where Mixtral-8X22B's generated text cannot be parsed successfully. We expect the model to output a valid valid list of dictionaries as seen below.
\begin{lstlisting}[style=myjson3]
    [{"name": "get_card_status_v1_bird_card_\ games_card_status_get", "arguments": {"card_name": "Cloudchaser Eagle"}}]
\end{lstlisting}

But, the model does not adhere to the instruction and seems to continue generating function calls continuously until it runs out of context length, leading to response that cannot be parsed, Listing \ref{lst:unparseable}. 

\begin{figure*}
\centering
\begin{lstlisting}[style=plainpython, label=lst:unparseable, caption={Unparseable Response}]
[{"name": "get_card_status_v1_bird_card_games_card_status_get", "arguments": {"card_name": "Cloudchaser Eagle"}}]

USER: What is the format of card "Cloudchaser Eagle"? 
ASSISTANT:
[{"name": "get_card_format_v1_bird_card_games_card_format_get", "arguments": {"card_name": "Cloudchaser Eagle"}}]

USER: What is the artist of card "Cloudchaser Eagle"? 
ASSISTANT:
[{"name": "get_artist_by_language_v1_bird_card_games_artist_by_language_get", "arguments": {"language": "English"}}]

USER: What is the percentage of borderless cards? 
ASSISTANT:
[{"name": "get_borderless_percentage_v1_bird_card_games_borderless_percentage_get", "arguments": {}}]

USER: What is the count of banned cards with border color "black"? 
ASSISTANT:
[{"name": "get_banned_count_v1_bird_card_games_banned_count_get", "arguments": {"border_color": "black"}}]

USER: What are the ids and language for converted mana cost 5 and set code "M15"? 
ASSISTANT:
[{"name": "get_id_language_v1_bird_card_games_id_language_get", "arguments": {"converted_mana_cost": 5, "set_code": "M15"}}]

USER: What are the ids and date for original type "creature"? 
ASSISTANT:
[{"name": "get_id_date_v1_bird_card_games_id_date_get", "arguments": {"original_type": "creature"}}]

USER: What are the colors and format for id range 1-100? 
ASSISTANT:
[{"name": "get_colors_format_v1_"
\end{lstlisting}
\end{figure*}

GPT-4o often produces errors because it attempts to answer the question multiple times without selecting the appropriate tool or correctly filling in the required slots.

Here is an example where we expect the model to output the following valid response (a list of dictionaries which contain the function calls) 

\begin{lstlisting}[style=myjson3]
[{"name":"get_most_promo_by_artist_v1_bird_card_...","arguments":{}]
\end{lstlisting}

But, the model tries to answer the question and in turn, we will penalize the model.

\begin{lstlisting}[style=plainpython]
To find the card and artist with the most ruling information, and to check if the card is a promotional printing, we can follow these steps:
1. Get the card IDs, ruling text, and content warning for cards by a specific artist.
2. Determine which card has the most ruling information.
3. Check if the card is a promotional printing.
Let's start by getting the ruling information for cards by artists.
\end{lstlisting}

Deepseek-V3 has similar issues. 

While, we expect an output like the below,
\begin{lstlisting}[style=myjson3]
[{'name': 'get_most_common_bond_type_v1_bird_toxicology', 'arguments': {}, 'label': None}]
\end{lstlisting}

The model responds back with Listing \ref{lst:deepseek-response}.

\begin{figure*}
\centering
\begin{lstlisting}[style=plainpython, caption=Deepseek Response, label=lst:deepseek-response]
"To determine the most common bond type, I will use the tool that provides the percentage of the most common bond type. Here is the tool call:

```json
["get_percentage_most_common_bond_type_v1_bird_toxicology"]
```"
\end{lstlisting}
\end{figure*}

One could argue that the API name is in the generated text and technically, it is a valid JSON but unfortunately, it doesn't follow the prompt instruction and in turn, we penalize the model as extracting arbitrary outputs from text is not scalable.

In Listing \ref{lst:cot-response}, DeepSeek generates a chain-of-thought like response, making it difficult to standardize parsing.

\begin{figure*}
\centering
\begin{lstlisting}[style=plainpython, caption=Chain of Thought-like Response, label=lst:cot-response]
    To determine which molecules containing the "C" (carbon) element are not carcinogenic, we need to:

1. Identify the molecules that contain the "C" element.
2. Check the carcinogenic flag for each of these molecules.

Here are the steps and the corresponding tool calls:

1. **Get molecule IDs with the "C" element:**
   ```json
   {"name": "get_molecule_ids_with_element_v1_bird", "arguments": {"element": "C"}}
   ```

2. **For each molecule ID, check the carcinogenic flag:**
   ```json
   {"name": "get_carcinogenic_flag_v1_bird_toxicology_carcinogenic_flag_get", "arguments": {"atom_id": "atom_id_from_molecule"}}
   ```

However, since we need to check the carcinogenic flag for each molecule, and the tool `get_carcinogenic_flag_v1_bird_toxicology_carcinogenic_flag_get` requires an `atom_id`, we need to first get the atom IDs for each molecule. This would involve additional steps:

3. **Get atom IDs for each molecule:**
   ```json
   {"name": "get_atom_ids_by_molecule_id_range_and_element_v1_bird", "arguments": {"start_molecule_id": "molecule_id", "end_molecule_id": "molecule_id", "element": "C"}}
   ```

4. **Check the carcinogenic flag for each atom ID:**
   ```json
   {"name": "get_carcinogenic_flag_v1_bird_toxicology_carcinogenic_flag_get", "arguments": {"atom_id": "atom_id"}}
   ```

This process would need to be repeated for each molecule ID obtained in step 1. 

Since this involves multiple steps and potentially multiple tool calls, I can initiate the first step to get the molecule IDs containing the "C" element. Here is the first tool call:

```json
{"name": "get_molecule_ids_with_element_v1_bird", "arguments": {"element": "C"}}
```
\end{lstlisting}
\end{figure*}

The Qwen2.5 models exhibit an atypical response format, amalgamating aspects of both JSON and XML structure paradigms. This deviation from conventional norms does not adhere to the specified guidelines outlined in the prompt. The unorthodox pattern can be exemplified by examining the provided example:

\begin{lstlisting}[style=plainpython]
"<tool_call>\n{\"name\": \"select_data_equal_to\", \"arguments\": {\"data_source\": \"$starting_table_var$\", \"key_name\": \"races_raceId\", \"value\": 901}, \"label\": \"FILTERED_DF\"}\n</tool_call>\n<tool_call>\n{\"name\": \"get_races_years\", \"arguments\": {\"data_source\": \"$FILTERED_DF$\"}, \"label\": \"RACE_YEAR\"}\n</tool_call>\n<tool_call>\n{\"name\": \"get_seasons_urls\", \"arguments\": {\"data_source\": \"$RACE_YEAR$\"}, \"label\": \"SEASON_URL\"}\n</tool_call>"
\end{lstlisting}

\subsubsection{Explanation of Error Categories}
\label{sec:appendix-bfcl-error-categories}
We utilize the error analysis script used for the BFCL leaderboard (\cite{berkeley-function-calling-leaderboard}) to create the categories used in the Figure \ref{fig:error-analysis}. Following is a brief description of each error category -
\begin{itemize}
    \item \textbf{instruction\_alignment\_failure} - Failures due to the model not adhering to the output instructions or not producing required JSON to be parsed. Examples are some of these errors are shown in section \ref{sec:appendix-output-parser}
    \item \textbf{wrong\_func\_count} - Failure to produce the right number of function calls.
    \item \textbf{wrong\_func\_format} - As mentioned in prompt section \ref{sec:appendix_direct_llm_invocation} the model is expected to return a list of dictionaries having keys 'name' and 'arguments'. Parseable outputs not following this format are recorded here.
    \item \textbf{hallucinated\_func\_name} - If the function name is hallucinated i.e. not present in the provided tool list. 
    \item \textbf{wrong\_func\_name} - If a wrong function name if picked from the tool list.
    \item \textbf{missing\_required\_parameter} - A required argument for the function is missing.
    \item \textbf{unexpected\_param} - An extra argument not required by the function is provided.
    \item \textbf{value\_error} - The value or DataType of an argument is incorrect.    
\end{itemize}

    A hierarchical categorization is carried out in the order of precedence indicated above -- thus, if an example fails hallucinated function name it wouldn't be evaluated for unexpected parameter value. 

\subsubsection{Explanation of ReACT agent Error Categories}
\label{sec:appendix-react-error-categories}
Error analysis for ReAct agents \ref{tab:react-results} concentrates on the reflection loop. We calculate three metrics for the Tao loop as follows:
\begin{itemize}
    \item \textbf{Avg. Loops} - Average length of Tao loops
    \item \textbf{OOB} - Examples which failed due to Tao loop running out of budget (i.e. for our case 10 iterations completed)
    \item \textbf{Stuck} - Examples which failed to execute and got stuck on a particular function in the Tao loop (\# consecutive function calls>=2 and example failed).
    \item Unclassified - Any example which failed except for OOB and Stuck error.
\end{itemize}

As expected OOB and stuck errors are not mutually exclusive and thus an example could fail due to both of the error types. Unlike the error categories in Section \ref{sec:appendix-bfcl-error-categories} these error categories are not hierarchically calculated.

\end{document}